\documentclass[a4paper,11pt]{article}
\pdfoutput=1 \usepackage{amssymb}
\usepackage{amsmath}
\usepackage{braket}
\usepackage{mathtools}
\usepackage[usenames,dvipsnames,svgnames,table]{xcolor}
\usepackage[utf8]{inputenc}
\usepackage{color}
\usepackage{cite}
\usepackage{subcaption}
\usepackage{lmodern}
\usepackage{footnote}
\usepackage[normalem]{ulem}
\usepackage{glossaries-extra}
\setabbreviationstyle[acronym]{long-short}
\glssetcategoryattribute{acronym}{nohyperfirst}{true}

\usepackage{slashed}
\usepackage[pdftex]{graphicx}
\usepackage{multirow}
\usepackage{jheppub}
\usepackage{here}
\usepackage{hyperref}
\usepackage{verbatim}
\usepackage[justification=centering,singlelinecheck=false]{caption}
\usepackage{cleveref}
\usepackage{listings}
\usepackage{fancyvrb}
\usepackage{dsfont}
\usepackage{nicefrac,xfrac}

\Crefname{equation}{eq.}{eqs.}
\Crefname{section}{section}{sections}
\Crefname{figure}{figure}{figures}
\Crefname{appendix}{appendix}{appendices}

\newcommand{\cA}{\mathcal{A}}
\newcommand{\cB}{\mathcal{B}}

\newcommand{\cD}[0]{\mathcal D}

\newcommand{\cK}[0]{\mathcal K}

\newcommand{\cM}[0]{\mathcal M}
\newcommand{\cO}[0]{\mathcal O}

\newcommand{\mc}[0]{\mathcal}

\newcommand{\df}[0]{\mathrm{df}}

\newcommand{\Kdf}[0]{{\cK_{\df,3}}}

\newcommand{\K}[0]{\mathcal K}
\newcommand{\kdf}{\mathcal{K}_{\text{df},3} }

\newcommand{\bm}[0]{\boldsymbol}

\newcommand{\astarlittle}[0]{\mbox{\small $\boldsymbol a^{\!*}$}}
\newcommand{\bstarlittle}[0]{\mbox{\small $\boldsymbol b^{\!*}$}}
\newcommand{\aprimestarlittle}[0]{\mbox{\small $\boldsymbol a'^{*}$}}
\newcommand{\bprimestarlittle}[0]{\mbox{\small $\boldsymbol b'^{*}$}}
\newcommand{\kprimelittle}[0]{\mbox{\small $\boldsymbol k'$}}
\newcommand{\aprimelittle}[0]{\mbox{\small $\boldsymbol a'$}}
\newcommand{\bprimelittle}[0]{\mbox{\small $\boldsymbol b'$}}
\newcommand{\alittle}[0]{\mbox {\small $\boldsymbol a$}}
\newcommand{\blittle}[0]{\mbox {\small $\boldsymbol b$}}
\newcommand{\klittle}[0]{\mbox {\small $\boldsymbol k$}}
\newcommand{\plittle}[0]{\mbox {\small $\boldsymbol p$}}

\newcommand{\Luscher}[0]{Luscher:1986n2,Luscher:1991n1}

\newcommand{\HH}[0]{Horz:2019rrn}
\newcommand{\HSQCa}[0]{Hansen:2014eka}
\newcommand{\HSQCb}[0]{Hansen:2015zga}
\newcommand{\LtoK}[0]{Hansen:2014eka}

\newcommand{\BHSQC}[0]{Briceno:2017tce}
\newcommand{\BHSnum}[0]{Briceno:2018mlh}

\newcommand{\dwave}[0]{Blanton:2019igq}

\newcommand{\largera}[0]{Romero-Lopez:2019qrt}

\newcommand{\isospin}[0]{Hansen:2020zhy}

\newcommand{\BSnondegen}[0]{Blanton:2020gmf}
\newcommand{\BStwoplusone}[0]{Blanton:2021mih}

\newcommand{\Akakia}[0]{Hammer:2017uqm}
\newcommand{\Akakib}[0]{Hammer:2017kms}

\newcommand{\ThreeQCDNumerics}[0]{%
Detmold:2011kw,
Mai:2018djl,
Horz:2019rrn,
Blanton:2019vdk,
Mai:2019fba,
Culver:2019vvu,
Fischer:2020jzp,
Hansen:2020otl,
NPLQCD:2020ozd,
Alexandru:2020xqf,
Brett:2021wyd,
Blanton:2021llb,
Mai:2021nul,
Draper:2023boj,
Garofalo:2022pux}
\newcommand{\ThreeBody}[0]{%
Detmold:2008gh,
Beane:2007qr,
Briceno:2012rv,
Polejaeva:2012ut,
Hansen:2014eka,
Hansen:2015zga,
Briceno:2017tce,
Hammer:2017uqm,
Konig:2017krd,
Hammer:2017kms,
Mai:2017bge,
Briceno:2018mlh,
Briceno:2018aml,
Blanton:2019igq,
Pang:2019dfe,
Jackura:2019bmu,
Briceno:2019muc,
Romero-Lopez:2019qrt,
Hansen:2020zhy,
Blanton:2020gha,
Blanton:2020jnm,
Pang:2020pkl,
Romero-Lopez:2020rdq,
Blanton:2020gmf,
Muller:2020vtt,
Blanton:2021mih,
Muller:2021uur,
Blanton:2021eyf,
Jackura:2022gib,
Muller:2022oyw}

\newcommand{\ThreeBodyReviews}[0]{%
Hansen:2019nir,
Rusetsky:2019gyk,
Mai:2021lwb,
Romero-Lopez:2021zdo,
Romero-Lopez:2022usb}

\newcommand{\spectral}[0]{%
Hansen:2017mnd,
Hansen:2019idp,
Bulava:2021fre,
Gambino:2022dvu,
DelDebbio:2022qgu,
Alexandrou:2022tyn}

\newacronym{CMF}{CMF}{center-of-momentum frame}

\DeclareFixedFont{\ttb}{T1}{txtt}{bx}{n}{9}
\DeclareFixedFont{\ttm}{T1}{txtt}{m}{n}{9}

\definecolor{deepblue}{rgb}{0,0,0.5}
\definecolor{deepred}{rgb}{0.6,0,0}
\definecolor{deepgreen}{rgb}{0,0.5,0}
\definecolor{jlab_red}{RGB}{192,39,45}
\definecolor{jlab_orange}{RGB}{249,102,0}
\definecolor{jlab_blue}{RGB}{47,122,121}
\definecolor{jlab_green}{RGB}{65,125,10}
\definecolor{jlab_blue}{RGB}{47,122,121}

\title{Three relativistic neutrons in a finite volume}

\author[a]{Zachary T. Draper}
\affiliation[a]{Physics Department, University of Washington, Seattle, WA 98195-1560, USA}

\author[b]{\!, Maxwell T. Hansen}
\affiliation[b]{School of Physics and Astronomy, University of Edinburgh, Edinburgh EH9 3JZ, UK}

\author[c]{\!, Fernando Romero-L\'opez}
\affiliation[c]{CTP, Massachusetts Institute of Technology, Cambridge, MA 02139, USA}

\author[a]{\!, and~Stephen~R.~Sharpe}

\emailAdd{ztd@uw.edu}
\emailAdd{maxwell.hansen@ed.ac.uk}
\emailAdd{fernando@mit.edu}
\emailAdd{srsharpe@uw.edu}

\abstract{We generalize the relativistic field-theoretic (RFT) three-particle finite-volume formalism to systems of three identical, massive, spin-$1/2$ fermions, such as three neutrons. This allows, in principle, for the determination of the three-neutron interaction from the finite-volume spectrum of three-neutron states, which can be obtained from lattice QCD calculations.}

\allowdisplaybreaks

\preprint{MIT-CTP/5539}
\begin{document}

\maketitle
\flushbottom
\clearpage

\section{Introduction}
\label{sec:intro}

First-principles lattice QCD (LQCD) calculations of two-baryon scattering amplitudes, including the properties of subthreshold bound states, are rapidly progressing (see, e.g., refs.~\cite{Beane:2006mx,Inoue:2011ai,Orginos:2015aya,Berkowitz:2015eaa,Yamazaki:2015asa,Wagman:2017tmp,Francis:2018qch,NPLQCD:2020lxg,Horz:2020zvv,Green:2021qol,Amarasinghe:2021lqa}). This includes the determination of masses and other properties of certain light nuclei and hypernuclei, albeit at heavier-than-physical quark masses~\cite{Beane:2009gs,NPLQCD:2011naw,NPLQCD:2012mex,Doi:2011gq,Yamazaki:2012hi}. However, less progress has been made using LQCD to determine the three-nucleon interaction, which plays an important role for nuclei near the neutron driplines, for nuclear saturation, and in determining the neutron-star equation of state~\cite{Hammer:2012id,Hebeler:2015hla}. To advance such calculations, a formalism relating finite-volume energies to infinite-volume forces must be developed.

Such relations often take the form of quantization conditions, which relate the finite-volume spectrum obtained from lattice calculations to infinite-volume scattering amplitudes. At this stage, the two-body formalism, based on the seminal work of L{\"u}scher~\cite{\Luscher}, is very well understood, and its application has become a standard technique in LQCD. The development of the three-body formalism is an active area of research, with many cases understood~\cite{\ThreeBody}, and with a rapidly growing number of applications to LQCD results~\cite{\ThreeQCDNumerics}. For recent reviews see refs.~\cite{\ThreeBodyReviews}. Other finite-volume approaches, which we do not discuss, include formulating EFTs in finite volume~\cite{Eliyahu:2019nkz,Detmold:2021oro,Meng:2021uhz,Sun:2022frr,Bazak:2022mjh}, the finite-volume Hamiltonian method (see, e.g., ref.~\cite{Wu:2014vma}), and the HAL-QCD method~\cite{Doi:2011gq}. Still another approach, which does not take advantage of finite-volume effects, is to determine multi-hadron observables via smeared spectral functions, extracted from a regulated estimate of the notoriously ill-posed inverse Laplace transform~\cite{\spectral}.

To date, the finite-volume three-body formalism has only considered spinless (scalar or pseudoscalar) particles, including both identical and non-identical particles and, in the latter case, non-degenerate masses. To study more interesting quantities in the realm of nucleon and nuclear physics---such as the Roper resonance, the tritium nucleus, or the three-nucleon force---one must understand how to include particles with spin in the formalism. In this work we take the first step in this direction by considering three identical spin-$1/2$ fermions. This allows us to address the issues arising from the presence of spin in the simplest context. It also gives access to a very important three-particle system, namely that of three neutrons. Further generalizations, e.g., to generic three-nucleon systems or the $N\pi+N\pi\pi$ system needed to describe the Roper resonance, will be left for future work.

We follow the relativistic field theory (RFT) approach, pioneered in refs.~\cite{\HSQCa,\HSQCb}. The main technical complication in the derivation is the description of spin in relativistic systems of three particles. This is mainly due to the fact that the projection of the spin along an axis is not Lorentz invariant. This leads to the mixing of spin-1/2 and spin-3/2 three-neutron states. The complication is absent in the finite-volume formalism for two particles with arbitrary spin~\cite{Briceno:2014oea}, since all pairwise interactions occur at fixed overall two-particle momentum. In addition, in two-nucleon systems, the total spin is effectively conserved, since spin-1 or spin-0 states have opposite parities and do not mix.

The formalism is only valid for center-of-momentum frame (CMF) energies up to the first inelastic threshold, which for the three-nucleon system occurs at the pion production threshold, $3N \to 3N+\pi$. For physical quark masses, this threshold occurs at energies for which the nucleons are in the nonrelativistic regime. Thus, one might argue that developing a relativistic formalism is overkill, as the NREFT approach of refs.~\cite{\Akakia,\Akakib} is simpler. However, we expect LQCD studies of three nucleons to use heavier-than-physical quarks in the near term, and in this case the inelastic threshold is higher, with relativistic effects correspondingly enhanced.\footnote{The relative magnitude of relativistic effects with respect to the leading nonrelativistic contribution can be quantified by the size of $\boldsymbol k^2/m_N^2$. In the vicinity of the $3m_N+m_\pi$ threshold, this quantity can take values up to 0.5 for a setting such as the one in Refs.~\cite{Horz:2020zvv,Amarasinghe:2021lqa}, with $m_\pi\simeq 700-800$ MeV.} In addition, as noted above, the complications introduced by spin must be understood in the relativistic domain to discuss higher baryon resonances such as the Roper.

As is always the case in the finite-volume three-particle relations, the formalism for three neutrons involves two steps: (1) deriving the three-neutron quantization condition, which connects the spectrum to an intermediate K-matrix, $\Kdf$; and (2) deriving an integral equation relating $\Kdf$ to the three-neutron scattering amplitude, $\cM_3$. We find that the resulting equations can be written in essentially identical form to those for identical scalars, with the complications from spin leading to additional indices, the presence of signs resulting from Fermi statistics, and the need for momentum-dependent spin transformation operators related to Wigner rotations.

A necessary step to analyze lattice QCD data using the quantization condition is to have parametrizations of the three-particle K-matrix, $\Kdf$. As in previous work~\cite{\HSQCa,\BHSnum,\dwave,\BStwoplusone}, we develop an expansion about threshold for the three-neutron amplitude. The constraints from Fermi statistics significantly restrict the allowed forms that can appear through next-to-leading order in this expansion.

The remainder of this paper is organized as follows. We begin, in \Cref{sec:recap}, recalling the essential features of the RFT formalism for identical scalars, from which the generalization presented here is developed. This section also introduces the required kinematic notation. The core of this work is \Cref{sec:QC}, in which we describe the derivation of formalism for three identical spin-$1/2$ particles. We divide this long section into three parts. In the first, \Cref{sec:relspin}, we describe the kinematic modifications that arise in the presence of spin, in particular due to the need for Wigner rotations. We then, in \Cref{sec:ModOfBB}, describe how the building blocks of the RFT formalism are modified in the presence of spin. Finally, in \Cref{sec:derivation}, we show how these new building blocks appear in the derivation, and discuss the generalization to all orders in the skeleton expansion. In order to make this paper more accessible, we present a separate summary of the final results in \Cref{sec:summary}. The parametrization of $\Kdf$ is described in \Cref{sec:Kdf}, and we conclude in \Cref{sec:conc}. We relegate technical details to three appendices. In \Cref{app:deriv} we describe the way in which the antisymmetry of three fermion fields alters the derivation, while in \Cref{app:symm} we detail the specific antisymmetrization of momenta needed to define the relation between $\Kdf$ and $\cM_3$. Finally, \Cref{app:twoderivs} provides additional technical details concerning the enumeration of operators appearing in \Cref{sec:Kdf}.

\section{Recap of the formalism for identical scalars\label{sec:recap}}

In this section we review the RFT finite-volume scattering formalism for identical spin-zero particles, derived in refs.~\cite{Hansen:2014eka,Hansen:2015zga}. The derivation assumes a $\mathbb Z_2$ symmetry that decouples the sectors with even and odd numbers of particles. For pions in isosymmetric QCD this symmetry is G-parity, and for nucleons in the energy range below pion production it is baryon number.

The formalism for spin-zero particles consists of two parts: the first is a quantization condition relating the finite-volume energy spectrum (in a periodic box with side-length $L$) to an intermediate quantity called $\mathcal K_{\df,3}$, and the second is a set of integral equations relating this quantity to the physical three-to-three scattering amplitude. Both parts can be derived using a particular finite-volume correlation function denoted $ \mathcal M_{3,L}^{(u,u)}$, which can be shown to satisfy the following relation:%
\footnote{%
In the derivation for three neutrons presented in \Cref{sec:derivation} below, we use a different, though closely related, correlator to derive the quantization condition, as it is conceptually simpler and has been used in previous work, e.g. refs.~\cite{\HSQCa,\isospin}. The connection to the three-particle scattering amplitude is, however, made using the three-neutron generalization of $ \mathcal M_{3,L}^{(u,u)}$.
}
\begin{align}
\label{eq:M3Ldef}
\mathcal M_{3,L}^{(u,u)}(P) & \equiv \mathcal D^{(u,u)} + \mathcal L^{(u)}_L \frac{1}{1 + \mathcal K_{\df,3} F_3} \mathcal K_{\df,3} \mathcal R^{(u)}_L \,.
\end{align}

All quantities here are matrices on an index space that parametrizes three degenerate particles of mass $m$ having fixed total energy $E$ and spatial momentum $\boldsymbol P = (2 \pi/L) \boldsymbol n_P$ in the finite-volume frame. As indicated, the value of $\boldsymbol P$ is constrained by the periodicity to be an integer three-vector ($\boldsymbol n_P \in \mathbb Z^3$) multiple of $2 \pi/L$. The energy and three-momentum are grouped into the four-vector $P^\mu = (E, \boldsymbol P)$ that appears in the argument of $\mathcal M_{3,L}^{(u,u)}$ on the left-hand side of \Cref{eq:M3Ldef}. With energy and momentum fixed, one can additionally select one of the three particles, called the spectator, and denote its on-shell four-momentum by $k^\mu = (\omega_{k}, \boldsymbol k)$ where $\omega_{k} = \sqrt{\boldsymbol k^2 + m^2}$. The remaining two particles, sometimes called the interacting pair or dimer, then have four-momentum $P^\mu - k^\mu$ and their squared energy in their two-particle CMF is given by
\begin{equation}
\label{eq:E2stark}
(P-k)^2 \equiv E_{2,k}^{* 2} = (E - \omega_k)^2 - (\boldsymbol P - \boldsymbol k)^2 \,.
\end{equation}
Since this energy is fixed by $E$, $\boldsymbol P$ and $\boldsymbol k$, the same is true for the magnitude of back-to-back momenta for the two dimer particles. As a result, the only remaining degree of freedom is directional, and can be decomposed in spherical harmonics with indices $\ell m$. Combining the spherical harmonic indices with the momentum $\boldsymbol k = (2 \pi/L) \boldsymbol n$ gives a discrete index space abbreviated as $\{k\ell m\}$. This is the space on which all matrices in \Cref{eq:M3Ldef} are defined.%
\footnote{%
Since we are considering identical particles, only even values of $\ell$ contribute. This will not be the case for three neutrons, due to the spin degree of freedom.
}

$\mathcal M_{3,L}^{(u,u)}$ depends on $\mathcal K_{\df,3}$, as shown, together with four additional matrices ($F_3$, $\mathcal D^{(u,u)} $, $ \mathcal L^{(u)}_L $ and $ \mathcal R^{(u)}_L$) that can each be expressed in terms of four more fundamental building blocks. The first of these is the single-particle energy, $\omega_k$, promoted to a matrix in a trivial way:
\begin{equation}
\omega_{\boldsymbol k' \ell' m', \boldsymbol k \ell m} \equiv \delta_{\boldsymbol k' \boldsymbol k} \delta_{\ell' \ell} \delta_{m' m} \sqrt{\boldsymbol k^2 + m^2} \,.
\end{equation}
The second building block is $\mathcal K_2(E, \boldsymbol P)$, which encodes the two-to-two scattering that arises as a subprocess:
\begin{align}
\label{eq:manyKs}
\mathcal K_{2, \boldsymbol k' \ell' m', \boldsymbol k \ell m}(E, \boldsymbol P) & = \delta_{\boldsymbol k' \boldsymbol k} \mathcal K_{2, \ell' m', \ell m}( E_{2,k}^*) = \delta_{\boldsymbol k' \boldsymbol k} \delta_{\ell' \ell} \delta_{m'm} \mathcal K_2^{(\ell)}( E_{2,k}^*) \,, \\[3pt]
\mathcal K_2^{(\ell)}(E_{2,k}^*)^{-1} & = \frac{1}{16 \pi E_{2,k}^*} \Big [ p \cot \delta^{(\ell)}(p) + \vert p \vert \big (1 - H(\boldsymbol k) \big ) \Big ] \bigg \vert_{p = q_{2,k}^*} \,,
\label{eq:Kell}
\end{align}
where $\delta^{(\ell)}(p)$ is the scattering phase shift and
\begin{equation}
\label{eq:q2kstar}
q^*_{2,k} = \sqrt{E_{2,k}^{*2}/4 - m^2}
\end{equation}
is the momentum of each of the particles in the dimer in their CMF. We have also introduced $H(\boldsymbol k)$, a smooth cutoff function defined, for example, in eqs.~(28)-(29) of ref.~\cite{Hansen:2014eka}. This is a non-standard part of the K-matrix definition used in this formalism, but is required to avoid inducing unwanted power-like volume effects, as explained in refs.~\cite{Hansen:2014eka,Hansen:2015zga}.

The final two building blocks are two finite-volume quantities denoted $F(E, \boldsymbol P, L)$ and $G(E, \boldsymbol P, L)$. These are defined, respectively, in eqs.~(22-24) of ref.~\cite{\HSQCa} and eq.~(B.3) of ref.~\cite{\isospin}.%
\footnote{%
Alternatively, $F(E, \boldsymbol P, L)$ can be obtained from $\textbf F^{\sf lab}$ in \Cref{eq:Flab} below by multiplying by $2\omega_k L^3/i$ and dropping spin indices, while $G(E,\boldsymbol P, L)$ can be obtained from $\textbf G^{\sf lab}$ in \Cref{eq:Glab} by multiplying by $-2\omega_p L^3/i$, and dropping the spin indices.
}
We give their respective extensions to spin one-half particles in \Cref{eq:Fspin,eq:Flab} (for $F$) and \Cref{eq:Gspin,eq:Glab} (for $G$) below. With these in hand $F_3$, for example, can be written as
\begin{equation}
\label{eq:F3def}
F_3 \equiv \frac{1}{2 \omega L^3} \bigg [ \frac{F}{3} - F \frac{1}{1 + \mathcal M_{2,L} G } \mathcal M_{2,L} F \bigg ] \,, \ \ \ \ \ \ \mathcal M_{2,L} \equiv \frac{1}{\mathcal K_2^{-1} + F} \,.
\end{equation}

The first application of $ \mathcal M_{3,L}^{(u,u)}$ is that, up to neglected $e^{- m L}$ effects, the three-particle finite-volume energies correspond to poles in the factor appearing between $ \mathcal L^{(u)}_L$ and $\mathcal R^{(u)}_L$ in \Cref{eq:M3Ldef}. This leads to the quantization condition
\begin{equation}
\label{eq:identQC}
\text{det}_{k \ell m} \big [1 + \mathcal K_{\df,3}(E_3^*) F_3(E_3, \boldsymbol P, L) \big ] =0 \,,
\end{equation}
valid for $m < E_3^* < 5m$, where we have labeled the energy coordinate by $E_3$ to emphasize that the result applies for finite-volume energies with three-particle quantum numbers, and $E_3^* = \sqrt{E_3^2 - \boldsymbol P^2}$ is the energy in the overall CMF.

Superficially, this looks very similar to the quantization condition for two-particle states, arising from poles in $\mathcal M_{2,L}$ with fixed values of the spectator momentum $\boldsymbol k = \boldsymbol k'$. In this case the quantization condition is defined over angular-momentum indices,
\begin{equation}
\text{det}_{\ell m} \big [1 + \mathcal K_{2}(E_2^*) F_2(E_2, \boldsymbol P_2, L) \big ] =0 \,,
\end{equation}
and is valid for $0 < E_2^* < 4 m$, where $(E_2, \boldsymbol P_2)=(E_3-\omega_k,\boldsymbol P-\boldsymbol k)$ is the two-particle total energy-momentum, and $E_2^*=\sqrt{E_2^2-\boldsymbol P_2^2}$ is the two-particle CMF energy. In this result, $ \mathcal K_{2}(E_2^*) = \mathcal K_{2, \ell' m', \ell m}(E_2^*)$ refers to the quantity in the middle equality of \Cref{eq:manyKs}, i.e.~the K matrix carrying only angular momentum indices, and
\begin{align}
F_{2, \ell' m',\ell m}(E_2, \boldsymbol P_2, L) & = F_{\boldsymbol k \ell' m', \boldsymbol k \ell m}(E_2 + \omega_k, \boldsymbol P_2 + \boldsymbol k, L) \,.
\end{align}

To implement the formalism in practice, one must truncate the sum over angular momentum at some $\ell=\ell_{\sf max}$ in order that the matrices are finite. Calculations to date have used $\ell_{\sf max}=0$, $1$ or $2$. For both the two- and three-particle quantization conditions, solving for the roots at fixed $L$ and $\boldsymbol P$ for a given parametrization of $ \mathcal K_{2}(E_2^*)$ and $ \mathcal K_{\df,3}(E_3^*) $ gives a prediction for the finite-volume energies. Conversely, extracting the energies from a numerical lattice calculation gives constraints on the two K-matrices and fitting to a range of parametrizations allows one to determine these infinite-volume quantities.

It now remains to relate the K matrices to the physical three-particle amplitude. This is done by using $ \mathcal M_{3,L}^{(u,u)}$ to derive an integral equation. To do so requires explicit definitions for the remaining quantities appearing in \Cref{eq:M3Ldef}:
\begin{align}
\label{eq:Ddef_scalar}
\mathcal D^{(u,u)} & \equiv - \frac{1}{1 + \mathcal M_{2,L} G } \mathcal M_{2,L} G \mathcal M_{2,L} (2 \omega L^3) \,, \\
\label{eq:Ldef_scalar}
\mathcal L^{(u)}_L & \equiv \bigg[ \frac{F}{2 \omega L^3} \bigg ]^{-1} F_3 \,, \\
\mathcal R^{(u)}_L & \equiv F_3 \bigg[ \frac{F}{2 \omega L^3} \bigg ]^{-1}\,.
\label{eq:Rdef_scalar}
\end{align}

With these in hand, the derivation of the integral equations follows in two steps. First one notes that in definition of $ \mathcal M_{3,L}^{(u,u)}$ the external kinematics are treated asymmetrically, as indicated by the $(u,u)$ superscript. As discussed in refs.~\cite{Hansen:2014eka,Hansen:2015zga}, and recalled in \Cref{app:symm}, this is corrected by a symmetrizing step in which one combines the matrix with appropriate spherical harmonics to obtain a function of the external momenta, and then sums over separate permutations of the initial and final momenta. We denote this combination by
\begin{align}
\mathcal M_{3,L}(P) &\equiv \mathcal S \Big [ \mathcal M_{3,L}^{(u,u)}(P) \Big ] \,.
\label{eq:symmetrizeM3}
\end{align}
The quantity on the left-hand side is a symmetric function of the three incoming and three outgoing momenta,
and thus does not have the $(u,u)$ superscript.

The second step is to formally take an ordered double limit in which $ \mathcal M_{3,L}(P) $ becomes the physical scattering amplitude
\begin{equation}
\label{eq:MLinfty}
\mathcal M_3(E_3^*) = \lim_{\epsilon \to 0^+} \lim_{L \to \infty} \mathcal M_{3,L}(E + i \epsilon, \boldsymbol P) \,.
\end{equation}
In this definition, the energy is made slightly complex so that the $L \to \infty$ limit is well-defined. The limit can be directly applied to the known functions $F$ and $G$ and, after sending $\epsilon \to 0$, this leads to integral equations relating $\mathcal K_2$ and $\mathcal K_{\df,3}$ to the scattering amplitude. These integral equations are given explicitly in ref.~\cite{Hansen:2015zga}, and solved in certain approximations in refs.~\cite{Hansen:2020otl,Jackura:2020bsk,Garofalo:2022pux,Dawid:2023jrj}.
We give the explicit forms of the integral equations for the three-neutron system in
\Cref{sec:inteqs} below.

\section{Formalism for three spin-half particles\label{sec:QC}}

In the following subsections, we present the extension of the formalism reviewed above to three identical spin-1/2 particles. The section is divided into four subsections. First, in \Cref{sec:relspin}, we introduce the different bases of spin states required to derive and express our final result. Then, in \Cref{sec:ModOfBB}, we work through each of the building blocks mentioned for spin-0 particles above, and explain how these quantities are modified for the case of spin-1/2 particles. This allows us to efficiently summarize the new derivation, first, in \Cref{sec:derivation}, for the simplest contributing classes of diagrams, and then, in \Cref{sec:allorders}, extending to all orders. The final result in then presented in the following section, \Cref{sec:summary}.

\subsection{Three-body spin states in a relativistic formalism\label{sec:relspin}}

We begin with a discussion of the construction of spin states in a relativistic framework, and the transformation properties of such states under Lorentz boosts. An extended discussion on this can be found, e.g.,~in ref.~\cite{Chung:1971ri}.

We start by considering the states of a particle at rest, with mass $m$, spin $s$ and azimuthal component $m_s$. The states are denoted by $\ket{s\, m_s}$ where, as usual, we take $\hat z$ as the quantization axis. These states transform in the usual way under a general rotation $R$,
\begin{equation}
U(R)\ket{s\, m_s} = \ket{s\, m'_s} \cD^{(s)}_{m'_s, m_s}(R)\,,
\end{equation}
where $U(R)$ is the unitary operator corresponding to $R$, $\cD^{(s)}_{m'_s, m_s}(R)$ is the Wigner matrix for angular momentum $s$, and there is an implicit sum over $m_s'$. We then construct states with nonzero momentum $\boldsymbol p$ as
\begin{equation}
\ket{\boldsymbol p, s\,m_s } = U(L( {\boldsymbol \beta_p})) \ket{s\, m_s},
\label{eq:spinstate}
\end{equation}
where $L( {\boldsymbol \beta}_p)$ is a pure boost with velocity ${\boldsymbol \beta}_p = \boldsymbol p / \omega_p$, with $\omega_p = \sqrt{\boldsymbol p^2 + m^2}$, such that the boosted state has four-momentum $p^\mu=(\omega_p,\boldsymbol p)$. In this way, $\ket{\boldsymbol p, s\,m_s }$ is defined unambiguously without specifying a quantization axis in the moving frame. This is referred to as the standard basis and corresponds in the spin-$1/2$ case precisely to the state with the spinor $u(\bm p, m_s)$.

The boost in \Cref{eq:spinstate} can be represented as
\begin{equation}
L( {\boldsymbol \beta}_p) = R(\theta_p, \hat {\boldsymbol n}_{p} ) \cdot L( \beta_p \hat {\boldsymbol z}) \cdot R(\theta_p, \hat {\boldsymbol n}_{p} )^{-1},
\end{equation}
where $R(\theta_p, \hat {\boldsymbol n}_{p} ) $ is a rotation that takes a vector in the $z$-direction to point along the indicated momentum, i.e.~$\hat {\boldsymbol p} = R(\theta_p, \hat {\boldsymbol n}_{p} ) \hat {\boldsymbol z}$. Here $\hat {\boldsymbol n}_{p}$ defines the axis about which the rotation is performed and $\theta_p$ is the rotation angle. Using this representation, it is straightforward to determine the transformation of the states in \Cref{eq:spinstate} under rotations to be
\begin{equation}
U(R) \ket{\boldsymbol p, s\,m_s } = \ket{R \boldsymbol p, s\,m'_s } \cD^{(s)}_{m'_s, m_s}(R)\,.
\label{eq:standardrot}
\end{equation}
This is advantageous as it is equivalent to the transformations of a nonrelativistic spin state.

The transformation of the states in \Cref{eq:spinstate} under boosts requires some additional discussion. This is due to the fact that the azimuthal component of the spin does not remain unchanged under a generic boost. Consider the transformation
\begin{equation}
U( L( {\boldsymbol \beta}_k) ) \ket{\boldsymbol p, s\,m_s } = U( L( {\boldsymbol \beta}_k) ) U(L( {\boldsymbol \beta_p})) \ket{s\, m_s},
\end{equation}
with ${\boldsymbol \beta}_k$ a generic velocity. To work this out, we can use the well-known result that the product of two boosts is equal to the combination of a single boost and a rotation,
\begin{equation}
\label{eq:doubleboost}
L( {\boldsymbol \beta_k}) L( {\boldsymbol \beta_p}) = L( {\boldsymbol \beta'}) R(\theta, \hat{\boldsymbol n}).
\end{equation}
Here ${\boldsymbol \beta'}$ represents the velocity of the resulting boost, while $\hat {\boldsymbol n}$ and $\theta$ are, respectively, the axis and angle of the required rotation, whose expressions we provide below. It follows that
\begin{equation}
U( L( {\boldsymbol \beta}_k) ) \ket{\boldsymbol p, s\,m_s } =
\ket{\boldsymbol p_k, s\,m'_s } \cD^{(s)}_{m'_s, m_s}(R(\theta, \hat{\boldsymbol n})) ,
\end{equation}
where $\boldsymbol p_k$ is the spatial component of the four-momentum resulting from boosting $p^\mu = (\omega_p, \boldsymbol p)$ with velocity ${\boldsymbol \beta}_k$. Since we only consider spin-1/2 particles in this work, we can represent the Wigner matrices as
\begin{equation}
\cD(R(\theta, \hat{\boldsymbol n})) = \mathds{1}
\cos\frac{\theta}{2} \, - i \boldsymbol \sigma \cdot \boldsymbol n \sin \frac{\theta}{2} \,,
\end{equation}
with $\boldsymbol \sigma$ a vector composed of the Pauli matrices. Here, and for the remainder of the main text, we suppress the $(s) = (1/2)$ superscript on the Wigner matrices.

It remains to provide the expressions for the axis and angle of the rotation
$R(\theta, \hat{\boldsymbol n})$,
\begin{equation}
\hat {\boldsymbol n} = \frac{\boldsymbol \beta_k \times \boldsymbol \beta_p }{|\boldsymbol \beta_k \times \boldsymbol \beta_p |},\qquad
\cos \theta = \frac{(1+\gamma_p + \gamma_k +\gamma')^2}{(1+\gamma_p)(1+\gamma_k)(1+\gamma')}-1,
\label{eq:ntheta}
\end{equation}
where $\gamma_k = (1-\beta_k^2)^{-1/2}$ and similarly for $\gamma_p$, and
\begin{equation}
\gamma' = \gamma_k \gamma_p (1 + \boldsymbol \beta_p \cdot \boldsymbol \beta_k ),
\label{eq:gammap}
\end{equation}
is the Lorentz factor of the boost $L({\boldsymbol \beta'})$. We also stress that, with these definitions, ${\sin\theta \ge 0}$. A result that will be useful later is that if $ \boldsymbol \beta_p$ and $ \boldsymbol \beta_k$ lie on the same axis, or if one or both vanish, then $\cos \theta =1$, and the Wigner matrix is just the identity.

\bigskip

With these identities in hand, we are now in position to define two coordinate systems that will play an important role in the derivation that follows and in our statement of the final result for the generalized quantization condition. As noted in the previous section, on-shell three-particle variables are abbreviated by $\{k\ell m\}$, where $k$ is shorthand for the spectator momentum $\bm k$, and $\ell m$ refer to the decomposition of the remaining pair into spherical harmonics {\em in their CMF}. Here we have the complication of an additional spin degree of freedom for each of the particles. Two natural choices arise for the definition of the quantization axes used to define spin degrees of freedom:
\begin{enumerate}
\item
The \emph{lab-frame axis}: Label the on-shell four-momenta in the finite-volume frame (also called the lab frame) as $k$, $a$ and $b$, so that $k+a+b=P$. Define the spin degree of freedom for all three particles via the boost of \Cref{eq:spinstate} directly to this frame, i.e. use the states $\ket{\bm k,m_{s}(\klittle) }\otimes \ket{\bm a, m_{s}(\alittle)}\otimes \ket{\bm b, m_{s}(\blittle) }$, where, for example,
\begin{equation}
\ket{\bm a,m_{s}(\alittle) } \equiv U(L(\boldsymbol \beta_a)) \ket{\bm 0, m_{s} } .
\label{eq:labframestate}
\end{equation}
The notation of matching the momentum label on the state $\bm a$ with that inside $m_s(\alittle)$ indicates that a single boost is applied starting in the particle's rest frame. As we will see below, this choice of quantization axis occurs most naturally for $\Kdf$.

\item
The \emph{dimer-frame} axis: For $k$, continue to define the spin degree of freedom in the finite-volume frame, i.e. to use the state $\ket{\bm k,m_{s}(\klittle) }$. For $a$ and $b$, however, boost the four-momenta to the two-particle CMF. In this frame the four-momenta are denoted $a^*$ and $b^*$, with $|\bm a^*| = |\bm b^*| = q_{2,k}^*$ fixed by the choices of $k$ and $P$. Define the spin degree of freedom for the $\boldsymbol a^*$ state via \Cref{eq:spinstate},
\begin{equation}
\ket{\bm a^*, m_{s}(\astarlittle)} = U(L(\boldsymbol \beta_{a^*})) \vert \bm 0, m_{s} \rangle \,,
\label{eq:dimerframestate}
\end{equation}
and define $ \ket{\bm b^*, m_{s}(\bstarlittle)} $ analogously. We stress that the notation matches that introduced above, in that the use of a single boost is represented by the presence of $\bm a^*$ both in the label of the state and as the argument of $m_s$. The reason for using these states is that the spin indices rotate like NR spinors in the two-particle CMF, as shown in \Cref{eq:standardrot}, and are thus naturally combined with the pair orbital angular momentum $\ell$. Altogether, the \emph{dimer-frame} axis convention uses spin indices $m_{s}(\klittle), m_{s}(\astarlittle), m_{s}(\bstarlittle)$. This choice occurs naturally for $\K_2$, and for the kinematic function $F$.
\end{enumerate}

To relate these two choices we need to boost the lab-frame axis states for the dimer to their CMF,
\begin{equation}
\ket{\bm a^*, m_{s}(\alittle)} \equiv U(L(-\boldsymbol \beta_{P-k})) \ket{\bm a, m_{s}(\alittle)} \,,
\end{equation}
and similarly for $\ket{\bm b^*,m_s(\blittle)}$, where $\boldsymbol \beta_{P - k} = (\boldsymbol P - \boldsymbol k)/(E - \omega_{\boldsymbol k})$. Here we anticipate that the spin indices after boosting do not match those of the state $ \ket{\bm a^*, m_{s}(\astarlittle)}$ by using $\bm a$, rather than $\bm a^*$, as the argument of $m_s$ in $\ket{\bm a^*, m_{s}(\alittle)}$. The relation between spin indices is obtained using
\begin{align}
\ket{\bm a^*, m_{s}(\alittle)} & = U(L(-\boldsymbol \beta_{P-k})) U(L(\boldsymbol \beta_{a})) \vert \bm 0, m_{s} \rangle \,, \\
& = U(L(\boldsymbol \beta_{a^*})) U(R_a) \vert \boldsymbol 0, m_{s} \rangle \,, \\
& = \ket{\bm a^*, m'_{s}(\astarlittle)} \mathcal D(R_a)_{m'_{s} m_{s}} \,.
\label{eq:basisswitch}
\end{align}
To obtain the first line we have used the definition of $ \ket{\bm a, m_{s}(\alittle)} $. We have then used the relation between boosts and rotations, \Cref{eq:doubleboost}, to express $ \ket{\bm a^*, m_{s}(\alittle)} $ as a rotation of $ \ket{\bm a^*, m_{s}(\astarlittle)} $. Here $R_a = R(\theta_a, \hat {\boldsymbol n}_a)$ is defined via \Cref{eq:ntheta,eq:gammap} with the replacements
\begin{equation}
\label{eq:ra_def}
\boldsymbol \beta_p \to \boldsymbol \beta_{a}= \frac{\bm a}{\omega_a}\,,
\ \ {\rm and}\ \
\boldsymbol \beta_k \to - \boldsymbol \beta_{P-k} = - \frac{\bm P-\bm k}{E-\omega_k} \,.
\end{equation}
Detailed expressions for $R_a$, and the corresponding rotation for $b$, are given below.

For both of the coordinate systems summarized above, the final step is to project the dimer-pair to definite orbital angular momentum, always using the quantization axis in the two-particle CMF. As a result, the orbital-angular momentum and spin axes are aligned in the dimer-frame convention, but not in the lab-frame convention. In addition, as discussed below in \Cref{sec:K2sub,sec:changeofbasis}, a better choice is to project onto pair total spin (0 or 1 for two spin-half particles). This, together with the antisymmetry of the three-neutron state, and parity, decouples the even and odd sectors of orbital angular momentum for infinite-volume quantities.

\subsection{Modifications to the RFT building blocks\label{sec:ModOfBB}}

We now summarize the modifications to the various quantities entering the generalized RFT formalism. As we proceed, we provide some intuition regarding these changes, leaving a more complete explanation to the derivation section below, \Cref{sec:derivation}.

\subsubsection{\texorpdfstring{$\Kdf$}{The divergence-free K-matrix}\label{sec:Kdfsub}}

As we describe in \Cref{sec:Kdf}, the three-particle K-matrix is most naturally expressed using spin components defined with respect to the lab-frame axis, since the spin components correspond to those of Dirac spinors. In particular, we are led to consider
\begin{equation}
\label{eq:all_arguments}
\left [ \mathcal K_\text{df,3}^{\sf{lab}} \right ]_{\bm m_s', \bm m_s} =
\mathcal K_\text{df,3}^{\sf{lab}} \big (\bm k', m_{s}({\klittle}') ;
\bm a', m_{s}({\alittle}'); \bm b', m_{s}({\blittle}') \big \vert
\bm k, m_{s}({\klittle}); \bm a, m_{s}({\alittle});
\bm b, m_{s}({\blittle}) \ \big )\,,
\end{equation}
where, on the left-hand side, we have collected the spin labels as subscripts, using the shorthand
\begin{equation}
\label{eq:spin_short}
\bm m'_s = \big (m_{s}({\klittle}') , m_{s}({\alittle}'), m_{s}({\blittle}') \big ) \,,
\qquad \ \ \,
\bm m_s = \big (m_{s}({\klittle}) , m_{s}({\alittle}), m_{s}({\blittle}) \big ) \,.
\end{equation}

For the quantization condition, however, it turns out to be most convenient to use spin components defined with the dimer-frame axis, as we discuss below. In terms of explicit coordinates, this is simply given by
\begin{multline}
\left[\Kdf \right]_{ \bm m'^*_s , \bm m^*_s} =
\Kdf\big ( \bm k', m_{s}({\klittle}') ; \bm a', m_{s}({\alittle'^*});
\bm b', m_{s}({\blittle'^*}) \big \vert \bm k, m_{s}({\klittle});
\bm a, m_{s}({\alittle^*}); \bm b, m_{s}({\blittle^*}) \big ) \,,
\label{eq:Kdf_dimerframeargs}
\end{multline}
where we have introduced a second set of abbreviated spin indices, analogous to \Cref{eq:spin_short} above, but defined in the dimer-axis frame:
\begin{equation}
\label{eq:dimeraxisms}
\bm m'^*_s =
\big (m_{s}({\klittle}') , m_{s}({\alittle'^*}), m_{s}({\blittle'^*}) \big ) \,, \qquad
\bm m^*_s =
\big (m_{s}(\klittle) , m_{s}(\astarlittle), m_{s}(\bstarlittle) \big ) \,.
\end{equation}

The relation between these two choices can be worked out by repeated use of \Cref{eq:basisswitch}. It is given by a rotation in spin space for fixed values of the momenta,
\begin{align}
\left[\Kdf \right]_{ \bm m'^*_s , \bm m^*_s} \equiv \mathcal D^{(k',a')\dagger}_{\bm m'^*_s \bm m''_s} \
\left[\mathcal K_{\text{df},3}^{\sf lab} \right]_{ \bm m''_s , \bm m'''_s}
\mathcal D^{(k,a)}_{\bm m'''_s \bm m^*_s} \,,
\label{eq:basisspinrotation}
\end{align}
where we have used shorthand for the change-of-basis matrices. For example, the right-most matrix is given by
\begin{equation}
\mathcal D^{(k,a)}_{\bm m'''_s \bm m^*_s}
=
\delta_{m'''_s({\boldsymbol k}) m_s({\boldsymbol k})}
\cD(R_a^{-1})_{m'''_{s}({\boldsymbol a}) m_{s}({\boldsymbol a^*})}
\cD(R_b^{-1})_{m'''_{s}({\boldsymbol b}) m_{s}({\boldsymbol b^*})}\,,
\label{eq:Dka}
\end{equation}
where the rotation $R_a$ is defined using \Cref{eq:ntheta,eq:gammap}, with the replacements of \Cref{eq:ra_def}. The explicit definition is $R_a = R(\theta, \hat {\boldsymbol n})$ with
\begin{align}
\begin{split}
\hat {\boldsymbol n} = -\frac{\boldsymbol \beta_{P-k} \times \boldsymbol \beta_{a} }
{|\boldsymbol \beta_{P-k} \times \boldsymbol \beta_{a} |},
\ \
& \quad \cos \theta = \frac{(1+\gamma_{a} + \gamma_{P-k} +\gamma')^2}
{(1+\gamma_{a})(1+\gamma_{P-k})(1+\gamma')}-1\,,
\ \ \\[5pt]
\gamma' &= \gamma_{P-k} \gamma_{a} (1 - \boldsymbol \beta_{P-k} \cdot \boldsymbol \beta_{a})\,.
\end{split}
\end{align}
$R_b$ is defined in the same way but with $\boldsymbol \beta_a \to \boldsymbol \beta_b$.

To convert this to the matrix entering the quantization condition, we apply the same steps that are used in the spin-zero formalism. The dimers (defined with momenta $\boldsymbol a$ and $\boldsymbol b = \boldsymbol P - \boldsymbol k - \boldsymbol a$ in the initial state and $\boldsymbol a'$ and $\boldsymbol b' = \boldsymbol P - \boldsymbol k' - \boldsymbol a'$ in the final state) are projected to definite orbital angular momentum in their two-particle CMFs. This is done via integration with a spherical harmonic and we express the action as a projection,
\begin{equation}
\mathcal P^{\hat{\boldsymbol a}^*}_{\ell m} \circ f(\boldsymbol k, \boldsymbol a, \boldsymbol b) = \frac{1}{4 \pi} \int d \Omega_{\hat{\boldsymbol a}^*} Y^*_{\ell m}(\hat{\boldsymbol a}^*) f^*(\boldsymbol k, \boldsymbol a^*) \,,
\label{eq:PWproj}
\end{equation}
for a generic function $f$. Here ${\boldsymbol a}^*$ is the spatial part of the four-momentum that results from boosting $a^\mu = (\omega_a, \boldsymbol a)$ with boost velocity $\boldsymbol \beta_k = - (\boldsymbol P - \boldsymbol k)/(E - \omega_k)$ and $f^*$ is just $f$ expressed in terms of the new coordinates as shown. (The dependence on $\boldsymbol b^* = - \boldsymbol a^*$ is redundant and is thus omitted within $f^*$.) We then define
\begin{equation}
\label{eq:dimer_kdf_matrix}
\left[\textbf K_{{\rm df,3}}\right]_{k'\ell' m' \bm m'^*_s; k \ell m \bm m^*_s} \equiv i [\mathcal P^{\hat{\boldsymbol a}'^*}_{\ell' m'}]^\dagger \circ \mathcal P^{\hat{\boldsymbol a}^*}_{\ell m} \circ \Kdf (\,\cdots) \,,
\end{equation}
where the $(\,\cdots)$ on the right-hand side indicates the full set of arguments shown in the first line of \Cref{eq:Kdf_dimerframeargs} above. The momenta $k' = \boldsymbol k'$ and $k = \boldsymbol k$ are expressed as indices on the left-hand side, and it is understood that these are in the finite-volume set. Here we have also introduced the boldface notation $\textbf K_{{\rm df,3}}$ to represent the matrix version of $\Kdf$ within the generalized spin formalism. The factor of $i$ on the right-hand side of \Cref{eq:dimer_kdf_matrix} is for convenience.%
\footnote{%
The use of boldface for matrix quantities entering the quantization condition follows the conventions of
ref.~\cite{\isospin}.
}

We now return to the issue of why the dimer-frame axis is the more natural for the quantization condition. The point is that we wish to be able to combine orbital and spin angular momenta of the dimer in a simple way, and this is only possible if both are defined with respect to the same frame. Since orbital angular momenta are defined in the dimer CMF, the spin variables need to be as well, and this corresponds to using the dimer-frame axis. We stress that the spin rotation in \Cref{eq:basisspinrotation} and the partial-wave projection in \Cref{eq:PWproj} do not commute, so there is not a simple relation between the matrix forms of $\kdf$ in the dimer- and lab-frame axes. In particular if the three-body K-matrix is parametrized in the lab frame, then the angular momentum projection of eq.~\eqref{eq:dimer_kdf_matrix} will include an integral over the angular dependence appearing inside the Wigner-D matrices.

\subsubsection{\texorpdfstring{$G$}{G}\label{sec:Gsub}}

Next we consider the kinematic switch function, denoted by $G$. The defining property of $G$ is that the spectator particle (and thus also the spectator momentum) differs between the left and right side of the insertion.

As illustrated in the central portion of \Cref{fig:deri}(c), $G$ is defined with an initial state containing the spectator momentum $k$ and the scattering pair (with momenta $p$ and $b_{pk} = P - p - k$), transitioning to a final state containing spectator $p$ and scattering pair (with momenta $k$ and $b_{pk}$). Henceforth, in this subsection we will abbreviate $b_{pk}$ by $b$, in order to avoid a proliferation of subscripts; this should not be confused with the momentum labeled $b$ in \Cref{fig:deri}(c). The exact definition makes use of momentum coordinates in the two-particle CMF of the scattering pair, both before and after the switch. Specifically we define $\boldsymbol{k}^*_p$ [$\boldsymbol{p}^*_k$] as the spatial part of the four-momentum arising from boosting $k^\mu = (\omega_k, \boldsymbol k)$ [$p^\mu = (\omega_p, \boldsymbol p)$] with boost velocity $-\boldsymbol \beta_{P-p}$ [$-\boldsymbol \beta_{P-k}$]. In addition to the momenta, we require the two-particle CMF energy $E_{2,k}^*$, defined in \Cref{eq:E2stark}, and the analogous quantity $E_{2,p}^*$ obtained by replacing $k$ with $p$. From these one can define the on-shell two-particle CMF momentum $q^*_{2,k}$ and $q^*_{2,p}$ using \Cref{eq:q2kstar}.

We are now ready to give the lab-frame axis version of $G$ for three spin-half particles:
\begin{multline}
[\textbf G^{\sf lab}]_{p\ell'm' \bm m'_s ; k\ell m \bm m_s} (E, {\boldsymbol P},L) \equiv
- \delta_{ m'_{s}({\boldsymbol p}), m_{s}({\boldsymbol p})}
\delta_{ m'_{s}({\boldsymbol k}), m_{s}({\boldsymbol k})}
\delta_{ m'_{s}({\boldsymbol b}), m_{s}({\boldsymbol b})}
\\ \times
\frac{i}{4 \omega_p \omega_k L^6} \frac{H(\boldsymbol{p}) H(\boldsymbol{k})}{b^2-m^2}
\frac{4\pi \mc{Y}_{\ell'm'}(\boldsymbol{k}^*_p) \mc{Y}^*_{\ell m}(\boldsymbol{p}^{*}_k) }{q_{2,p}^{*\ell'}\,q_{2,k}^{*\ell} } \,,
\label{eq:Glab}
\end{multline}
where $\mc{Y}_{\ell m}(\boldsymbol x) = \vert \boldsymbol x \vert^{\ell} Y_{\ell m}(\hat{\boldsymbol x})$
is a harmonic polynomial, and
\begin{equation}
b^2 = (E - \omega_k - \omega_p)^2 - (\boldsymbol P - \boldsymbol p - \boldsymbol k)^2 \,.
\end{equation}
The smooth cutoff function $H(\boldsymbol k)$ was already introduced in \Cref{eq:Kell}. In the context of three neutrons, we note that the allowed support for the $H$ function is set by the left-hand cut arising from $t$-channel pion exchange in the two-neutron amplitude. A similar situation is discussed in refs.~\cite{Blanton:2021eyf,Draper:2023boj} in the context of the RFT formalism for $K \pi \pi$ and $K K \pi$, see also ref.~\cite{Raposo:2023nex}. Relative to the spin-zero $G$ function discussed in \Cref{sec:recap} [see the explanation above \Cref{eq:F3def}], the definition (\ref{eq:Glab}) contains an extra factor of $i/(2 \omega_p L^3)$, matching the conventions of ref.~\cite{\isospin}.

There are two key new features relative to the form for spin-zero particles. The first is the overall minus sign. This results from the antisymmetry of the fermionic multi-particle state or, equivalently, from the anti-commutation of Grassmann variables in evaluating Feynman diagrams. This is discussed in more detail in \Cref{sec:contributingdiagrams} below.

The second new feature is the appearance of a product of Kronecker deltas in the spin components. This encodes the fact that, for spin components defined in the lab frame, $G$ simply acts like an identity matrix in spin space, a point that will be discussed further in \Cref{sec:finitevolumecorrelator}. Strictly speaking this only holds when all three particles are on shell, which is not the case in general: although $k$ and $p$ are on shell by construction, $b$ is not. This is potentially problematic because the lab-frame state $\ket{\bm b,m_{s}(\blittle) } $, given by \Cref{eq:labframestate}, is defined only for on-shell four-momenta. In particular, the boost velocity used to define the state is $\boldsymbol \beta_b\equiv \boldsymbol b/\omega_b$, even though $b^0 \ne \omega_b$ in general. The resolution to this issue is that all that matters for the derivation of the quantization condition is that the choice of state is correct on shell, i.e. at the pole where $b^2=m^2$. Choices that differ off shell lead to finite shifts in $\Kdf$, since the differences cancel the pole.

We stress that the Kronecker deltas in \Cref{eq:Glab} cannot be written as $\delta_{\bm m'_s \bm m_s}$, because the order of spin components in the compound labels does not match:
\begin{equation}
\bm m'_s = \big (m_{s}(\plittle) , m_{s}({\klittle}), m_{s}({\blittle}) \big ) \,,
\qquad \ \ \,
\bm m_s = \big (m_{s}({\klittle}) , m_{s}(\plittle), m_{s}({\blittle}) \big ) \,.
\end{equation}

Non-trivial spin structure arises when we transform to the dimer-axis frame. The transformation is similar to that for $\Kdf$ discussed in the previous subsection,
\begin{equation}
\textbf G_{p\ell' m' \bm m'^*_s; k \ell m \bm m^*_s}
=
\mathcal D^{(p,k)\dagger}_{\bm m'^*_s \bm m''_s} \
\textbf G^{\sf lab}_{p\ell' m' \bm m''_s; k \ell m \bm m'''_s} \
\mathcal D^{(k,p)}_{\bm m'''_s \bm m^*_s} \,,
\label{eq:Gspin}
\end{equation}
where $\bm m_s''$ and $\bm m'''_s$ are summed.

The change of basis matrices are given explicitly by
\begin{align}
\mathcal D^{(k,p)}_{\bm m'''_s \bm m^*_s} &
= \delta_{m'''_{s}({\boldsymbol k}) m_{s}({\boldsymbol k})}
\cD(R_p^{-1})_{m'''_{s}({\boldsymbol p}) m_{s}({{\boldsymbol p^*}})}
\cD(R_{b_k}^{-1})_{m'''_{s}({{\boldsymbol b_k}}) m_{s}({{\boldsymbol b_k^*}})}\,,
\\
\mathcal D^{(p,k)}_{\bm m'''_s \bm m^*_s} &
= \delta_{m'''_{s}({\boldsymbol p}) m_{s}({\boldsymbol p})}
\cD(R_k^{-1})_{m'''_{s}({\boldsymbol k}) m_{s}({{\boldsymbol k^*}})}
\cD(R_{b_p}^{-1})_{m'''_{s}({{\boldsymbol b_p}}) m_{s}({{\boldsymbol b_p^*}})}\,.
\end{align}
These depend on a total of four rotations, denoted by $R_p$, $R_{b_k}$, $R_k$ and $R_{b_p}$ as shown. Each of the four is induced by relating two successive boosts to a single boost, via \Cref{eq:doubleboost}. Thus all rotations can be expressed according to \Cref{eq:ntheta}. Specifically, we define
\begin{align}
R_p & : \qquad \hat {\boldsymbol n} = -\frac{\boldsymbol \beta_{P-k} \times \boldsymbol \beta_p }{|\boldsymbol \beta_{P-k} \times \boldsymbol \beta_p |},\qquad
\cos \theta = \frac{(1+\gamma_p + \gamma_{P-k} +\gamma'_{P-k,p})^2}{(1+\gamma_p)(1+\gamma_{P-k})(1+\gamma'_{P-k,p})}-1\,, \\
R_k & : \qquad \hat {\boldsymbol n} = - \frac{\boldsymbol \beta_{P-p} \times \boldsymbol \beta_k}{|\boldsymbol \beta_{P-p} \times \boldsymbol \beta_k |},\qquad
\cos \theta = \frac{(1+\gamma_k + \gamma_{P-p} +\gamma'_{P-p,k})^2}{(1+\gamma_k)(1+\gamma_{P-p})(1+\gamma'_{P-p,k})}-1\,,
\end{align}
where
\begin{align}
\gamma'_{P-k,p} & = \gamma_{P-k} \gamma_p (1 - \boldsymbol \beta_{P-k} \cdot \boldsymbol \beta_p ) \,,
&
\gamma'_{P-p,k} & = \gamma_{P-p} \gamma_k (1 - \boldsymbol \beta_{P-p} \cdot \boldsymbol \beta_k ) \,.
\end{align}
$R_{b_k}$ is defined as for $R_p$, but with $\boldsymbol \beta_p$ replaced by $\boldsymbol \beta_b=\bm b/\omega_b$, and $\gamma_p$ replaced by $\gamma_b=\sqrt{1/(1-\beta_b^2)}$. Similarly, $R_{b_p}$ is defined as for $R_k$, but with $\boldsymbol \beta_p$ replaced by $\boldsymbol \beta_b$ and $\gamma_k$ replaced by $\gamma_b$. As noted above, since $b$ is off shell in general, the choice of boost velocities is only unambiguous at the on-shell point. Any off-shell extension choice that is used consistently is sufficient to perform the derivation, and our choice here is to use $\boldsymbol \beta_b=\bm b/\omega_b$ rather than, say, $\bm b/b^0$.

\subsubsection{\texorpdfstring{$F$}{F}\label{sec:Fsub}}

We next turn to the kinematic function $F$, which implements the sum-integral difference for a loop involving two of the three-particles. The definition in the lab-axis frame is obtained from the standard form for $F$ for scalar particles by adding Kronecker deltas in spin space,
\begin{multline}
[\textbf F^{\sf lab}]_{k' \ell'm' \bm m'_s ; k\ell m \bm m_s} (E, {\boldsymbol P},L) \equiv
\delta_{\bm m'_s \bm m_s} \delta_{k'k} \frac{i H(\boldsymbol k)}{2 \omega_k L^3}
\frac12 \bigg [\frac{1}{L^3} \sum_{\boldsymbol a} - {\text{p.v.}} \int_{\boldsymbol a} \bigg ]
\\
\times \frac{4 \pi \mathcal Y_{\ell' m'}(\boldsymbol a^*_k)
\mathcal Y^*_{\ell m}(\boldsymbol a^*_k)}{2 \omega_a (b^2 - m^2) } \frac{1}{(q^*_{2,k})^{\ell+\ell'}} \,,
\label{eq:Flab}
\end{multline}
where here the order of the compound spin indices do match, such that we can use
\begin{equation}
\delta_{\bm m'_s \bm m_s} =
\delta_{ m'_{s}({\boldsymbol{k}}) m_{s}({\boldsymbol{k}})}
\delta_{ m'_{s}({\boldsymbol{a}}) m_{s}({\boldsymbol{a}})}
\delta_{ m'_{s}({\boldsymbol{b}}) m_{s}({\boldsymbol{b}})} \,.
\end{equation}
This definition of F differs from that for scalar particles, given in eqs.~(22-24) of ref.~\cite{\HSQCa}, by a factor of $i/(2 \omega_k L^3)$, as well as by the addition of the spin factors, and thus follows the normalization conventions of ref.~\cite{\isospin}.

The quantities in \Cref{eq:Flab} are defined as follows. The four-momentum $b$ is given by $b^\mu = (E - \omega_k - \omega_a, \boldsymbol P - \boldsymbol k - \boldsymbol a)$, while the on-shell magnitude $q^{*}_{2,k} $ is defined in \Cref{eq:q2kstar}. Following the usual pattern, $\boldsymbol a^*_k$ is the spatial part of the four-momentum resulting from boosting $a^\mu = (\omega_a, \boldsymbol a)$ with boost velocity $-\boldsymbol \beta_{P - k}$. The sum runs over values of $\boldsymbol a = 2 \pi \boldsymbol n_a/L$ where $\boldsymbol n_a$ is a three-vector of integers. The notation p.v. indicates a principal value pole prescription, including the possible extensions discussed in ref.~\cite{\largera}. Finally, it is understood that an ultraviolet cutoff must be included to evaluate the sum and integral separately. Any dependence on the cutoff vanishes in the difference as can be shown using the Poisson summation formula. The numerical evaluation of the sum-integral difference is discussed in more detail, e.g., in appendix B of ref.~\cite{Briceno:2018mlh} and also in appendix B of ref.~\cite{\dwave}.

As with $G$, this quantity must reflect the exchange properties of identical fermions. One aspect of this is the symmetry factor of $1/2$ for the $ab$ loop, which is present exactly as in the spin-zero case. To understand additional consequences, in particular of the antisymmetry, it is useful to transition from the lab-axis frame and dimer-axis frame. However, in this case there is no distinction
\begin{equation}
\textbf F
=
\textbf F^{\sf lab} \,.
\label{eq:Fspin}
\end{equation}
This is an important simplification as the change of basis matrices would in fact depend on the summed coordinate. However, since $\textbf F^{\sf lab}$ is diagonal in spectator momentum, the dimer frame is the same in the initial and final states. As a result the change of basis matrices exactly cancel.

\subsubsection{\texorpdfstring{$\mathcal K_2$}{The two-particle K-matrix}\label{sec:K2sub}}

The final building block is the two-particle K matrix. This quantity is naturally defined in the \emph{dimer-axis frame}, and we discuss only this version of the K matrix. It can be written in a manner analogous to that for scalars, \Cref{eq:manyKs},
\begin{equation}
\label{eq:boldK2_spin_def}
[\textbf K_2]_{k' \ell'm' \bm m'^*_s ; k\ell m \bm m^*_s} (E, {\boldsymbol P}) = i \delta_{k'k} 2 \omega_k L^3
\mathcal K_2^{(\ell' m' \bm m'^*_{s}, \ell m \bm m^*_{s} )}(E_{2,k}^*) \,,
\end{equation}
with $ \mathcal K_2^{(\,\cdots )}(E_{2,k}^*) $ on the right-hand side
a generalization of the quantity $ \mathcal K_2^{(\ell)}(E_2^*) $ in \Cref{eq:manyKs}
to the case of spin one-half particles, with the additional factor of $i 2 \omega_k L^3$
to match the convention in ref.~\cite{\isospin}. As above, the superscripts $\bm m^*_{s} $ and $\bm m'^*_{s}$ indicate that the spin quantization axis is defined in the two-particle CMF. The role of the spectator here is trivial and the K matrix can be unpacked as
\begin{equation}
\mathcal K_{2}^{(\ell' m' \bm m'^*_{s}, \ell m \bm m^*_{s} )}(E_{2,k}^*) =
\delta_{m'_s(\boldsymbol{k}) m_s(\boldsymbol{k})} \
\mathcal K_{2}^{[\ell' m' m'_{s}({{\boldsymbol a'^*}}) m'_{s}({{\boldsymbol b'^*}})] , \,
[\ell m m_{s}({{\boldsymbol a^*}}) m_{s}({{\boldsymbol b^*}}) ]}(E_{2,k}^*) \,.
\end{equation}
In words, the incoming state is labeled with orbital angular momentum $\ell, m$ together with spin components $m_{s}(\astarlittle) $ and $m_{s}(\bstarlittle)$, and the outgoing state carries the same set with primes as indicated.

To parametrize $\mathcal K_2$, it is more common to work in the basis which diagonalizes the total spin of the dimer, $s$. This can take the values $s=0$ (spin singlet) or $s=1$ (spin triplet). Enforcing antisymmetry, the singlet can only couple to even values of $\ell$, while the triplet couples only to odd values. Although not mentioned above, these constraints apply also to $\Kdf$, $F$ and $G$.

In general, $s$ and $\ell$ are not conserved by two-particle interactions, and one should convert to the basis in which the total dimer angular momentum, $j$, and its azimuthal component, $\mu_j$, are diagonal. One simplification here, however, is that there is no mixing between $s=0$ and $1$ states for two identical fermions, because their parities, given by $(-1)^\ell$, are opposite. Furthermore, if $s=0$ then $j=\ell$ and thus $\ell$ is also conserved (and takes only even values). On the other hand, if $s=1$, then $j=\ell+1,\,\ell,\, \ell-1$, so that, given that $\ell$ is odd, each odd value of $j$ arises from only a single value of $\ell$, while for all even values of $j$ except zero there are two choices of $\ell$.

We now describe in detail the conversions between bases. The first step is to convert to the basis of total dimer spin $s$ and azimuthal component $m$:
\begin{multline}
\mathcal K_{2}^{(s,\ell', m', \mu'_s,\ell, m ,\mu_s )} \delta_{s' s}
= \sum_{ m'_{s}(\boldsymbol a'^*) , m'_{s}(\boldsymbol b'^*) ,
m_{s}(\boldsymbol a^*) , m_{s}(\boldsymbol b^*) }
\Big \langle s' , \mu'_s \Big \vert \nicefrac12 \ m'_{s}(\aprimestarlittle) ,
\ \nicefrac12 \ m'_{s}(\bprimestarlittle) \Big \rangle
\\[5pt] \times
\mathcal K_{2}^{[\ell' m' m'_{s}(\boldsymbol a'^*) m'_{s}(\boldsymbol b'^*)] , \, [\ell m m_{s}({\boldsymbol a^*}) m_{s}({{\boldsymbol b^*}}) ]}
\Big \langle \nicefrac12 \ m_{s}(\astarlittle) , \ \nicefrac12 \ m_{s}(\bstarlittle)
\Big \vert s , \mu_s \Big \rangle
\,,
\label{eq:K2totals}
\end{multline}
where the angle-bracketed quantities are standard Clebsch-Gordan coefficients, and the conservation of $s$ is enforced by the Kronecker delta on the left-hand side. From here one can reach the final basis via
\begin{equation}
\mathcal K^{(j, s, \ell', \ell)}_{2} \ \delta_{j'j} \delta_{\mu'_j \mu_j}
= \sum_{\mu_s, \mu_s'} \big \langle j' , \mu'_j \vert \ell' m', s \mu'_s \big \rangle
\mathcal K_{2}^{(s , \ell', m' , \mu'_s, \ell, m, \mu_s )}
\big \langle \ell m, s \mu_s \vert j , \mu_j \big \rangle
\,.
\label{eq:K2totalj}
\end{equation}
Then one can write%
\footnote{%
This form assumes that the generalization of the p.v.~prescription to include an ``$I_{\rm PV}$ term'' is not being made. The generalization to this case can be easily determined from ref.~\cite{\largera}.
}
\begin{equation}
[\mathcal K_2(E_{2,k}^*)^{-1} ]^{(j, s, \ell', \ell)} = [\mathcal M_2(E_{2,k}^*)^{-1} ]^{(j, s, \ell', \ell)} + \delta_{\ell', \ell} \frac{ \vert p \vert \big (1 - H(\boldsymbol k) \big ) + i p }{16 \pi E_{2,k}^*} \bigg \vert_{p = q_{2,k}^*} \,,
\end{equation}
with subthreshold analytic continuation following from $- i p \to \vert p \vert$ for $p^2 < 0$. Here $\mathcal M_2(E_{2,k}^*)$ is the standard scattering amplitude.

The parametrization of the resulting scattering amplitudes depends on whether there is mixing between different values of $\ell$. The simplest case is $s=0$, which, as discussed above, always involves only a single channel with even $\ell=j$. Then we write the scattering amplitude in terms of a phase shift
\begin{equation}
[ {\mathcal M}_2(E_{2,k}^*)^{-1} ]^{(j, s=0, \ell', \ell)} + \delta_{\ell', \ell} \frac{ i p }{16 \pi E_{2,k}^*} \bigg \vert_{p = q_{2,k}^*} = \delta_{\ell', \ell} \frac{ q_{2,k}^* \cot \delta^{(j=\ell, s=0)}(q_{2,k}^*)}{16 \pi E_{2,k}^*} \,.
\end{equation}
This, in turn, can be parametrized with an effective range expansion
\begin{equation}
p^{2 \ell + 1} \cot \delta^{(j=\ell, s=0)}(p) = - \frac{1}{a_0} + \sum_{n=1}^\infty a_n (p^2)^n \,.
\end{equation}
Two-channel mixing is possible for $s=1$, but only for $j=2,4,6, \dots$, for which both $\ell=j\pm 1$ are possible. In such cases one can parametrize the phase shift using, for example, the Blatt-Biedenharn parametrization~\cite{PhysRev.86.399,deSwart:1995ui}. For $j=0$ and $1$, however, which arise from $\ell=1$, there is only a single channel and one can proceed as for $s=0$.

\subsection{Derivation\label{sec:derivation}}

In the previous section we have provided the generalization of the four key quantities required to develop the finite-volume formalism for three identical spin-$1/2$ particles. The generalized version for each of the four is written in bold face, in particular $\textbf K_{{\rm df,3}}$ [section~\ref{sec:Kdfsub}], $\textbf G$ [section~\ref{sec:Gsub}], $\textbf F$ [section~\ref{sec:Fsub}] and $\textbf K_2$ [section~\ref{sec:K2sub}]. In this section, we show how, using these building blocks, the derivation of refs.~\cite{\HSQCa,\HSQCb} can be extended to apply to identical spin-$1/2$-particle systems. For simplicity, we use the language of three neutrons in the following, but emphasize that the formalism we obtain holds for any system of three spin-$1/2$ particles.

The derivation proceeds in several steps. First, in \Cref{sec:finitevolumecorrelator}, we define an appropriate finite-volume correlator, the poles in which give the finite-volume energies. Next, in \Cref{sec:contributingdiagrams}, we analyze the three classes of Feynman diagrams that enter into the skeleton expansion of the correlator. Third, in \Cref{sec:allorders}, we explain how the results generalize to all orders. We also present the result for a related quantity, the finite-volume unsymmetrized scattering amplitude. These ingredients are then used in the following section, \Cref{sec:summary}, to determine the quantization condition and the relation between $\textbf K_{{\rm df,3}}$ and the infinite-volume scattering amplitude. This basic workflow is analogous to the presentation of ref.~\cite{\isospin}, in which the formalism of refs.~\cite{\HSQCa,\HSQCb} was generalized to describe all three-pion states with generic isospin.

\subsubsection{Finite-volume correlator\label{sec:finitevolumecorrelator}}

The Minkowski-space finite-volume correlation function is defined as
\begin{equation}
\label{eq:CLdef}
C_L(P) \equiv \int_L d^4x \, e^{i P \cdot x } \, \langle \text{T} \mathcal O(x) {\mathcal O}^\dagger(0) \rangle_L \,,
\end{equation}
where $P \cdot x = E x^0- \boldsymbol P \cdot \boldsymbol x $, and
\begin{equation}
\int_L d^4x \equiv \int dx^0 \int_{L} d^3 \boldsymbol x \,,
\end{equation}
is an integral over the infinite time direction and three finite spatial directions. The operators $\mathcal O^\dagger(0)$ and ${\mathcal O}(x)$ respectively create and annihilate three-neutron states. They are assumed to have support for a finite range of times around the nominal position of the operator, and can either be local or delocalized in space. In \Cref{eq:CLdef}, $\text{T}$ indicates time ordering, $P^\mu = (E, \boldsymbol P)$ is a four-vector containing the total energy and momentum, and the subscript $L$ indicates that the quantity is evaluated in a finite cubic spatial volume (with periodicity $L$ in each of the three spatial directions).

It will be useful for the following derivation to give explicit definitions of the operators $\mathcal O$ and ${\mathcal O}^\dagger$. Let $\mathcal N_\alpha(x)$ denote a single-nucleon spin-$1/2$ field with Dirac index $\alpha$. We first define the momentum-space operators
\begin{align}
\widetilde {\mathcal O}^\dagger_{r_1,r_2,r_3}(p_1,p_2,p_3) & = \prod_{i=1}^3
\left\{ u^{r_i}_{\alpha_i}(\boldsymbol p_{\alpha_i}) \int_L d^4 x_i \, e^{-i p_i \cdot x_i }
\overline{ \mathcal N}_{\alpha_i}(x_i) \right\} \,,
\label{eq:Otildedaggerdef}\\
\widetilde {\mathcal O}_{r_1,r_2,r_3}(p'_1,p'_2,p'_3) & = \prod_{i=1}^3
\left\{ \overline u^{r_i}_{\alpha_i}(\boldsymbol p'_{\alpha_i}) \int_L d^4 x_i \, e^{i p'_i \cdot x_i }
\mathcal N_{\alpha_i}(x_i) \right\}\,,
\label{eq:Otildedef}
\end{align}
where each field on the right-hand side is contracted with a spinor $\overline u^{r_i}_{\alpha_i}(\boldsymbol p'_{\alpha_i})$ or $u^{r_i}_{\alpha_i}(\boldsymbol p_{\alpha_i})$, with $r_i$ indicating a specific spin component. We then obtain $\cO(x)$ by integrating over momenta, weighted with a ``form factor'' that depends on the precise form of the operator. In particular, the Fourier transform of $\cO(x)$ is given by
\begin{equation}
\label{eq:Oxdef}
\int_L d^4x \, e^{i P \cdot x } \, \mathcal O(x) = \int_{p_1', \,L} \int_{p_2',\, L} \int_{p_3',\, L}
\, \sum_{r_1,r_2,r_3} \widetilde {\mathcal O}_{r_1,r_2,r_3}(p'_1,p'_2,p'_3)
\overline f_{r_1r_2r_3}(p_1',p_2',p_3') \,,
\end{equation}
where
\begin{equation}
\overline f_{r_1r_2r_3}(p_1',p_2',p_3') = 2\pi \delta(E - p'^{0}_1- p'^{0}_2- p'^{0}_3)\,
L^3 \delta_{\bm p_1'+\bm p_2'+\bm p_3', \bm P} \, f_{r_1r_2r_3}(p_1',p_2',p_3')\,.
\label{eq:fdef}
\end{equation}
In \Cref{eq:Oxdef}, we are using the finite-volume version of the momentum-space integral,
\begin{equation}
\int_{p,\,L} = \frac{1}{L^3} \sum_{\boldsymbol p} \int \! \frac{d p^0}{2 \pi} \,,
\end{equation}
in which the sum runs over all values of $\boldsymbol p = (2 \pi/L) \boldsymbol n$, where $\boldsymbol n$ is a three-vector of integers ($\boldsymbol n \in \mathbb Z^3$). We refer to this set of momenta as the ``finite-volume set.'' In \Cref{eq:fdef}, the first two factors on the right-hand side enforce four-momentum conservation, while $f$ is a smooth function of momenta that depends on the form of $\cO(x)$. The detailed form of $f$ is not important. However, its $p_i^0$ dependence must ensure that $\mathcal O(x)$ is localized in time, as described above.

Two technical points deserve mention. First, in \Cref{eq:CLdef}, the time ordering applies to the positions of all the single-particle operators contained in the composite operators $\cO$ and $\cO^\dagger$, and not just to the component $x_0$. In particular, all possible orderings of the $\mathcal N$ and $\overline {\mathcal N}$ fields arise in general, since these are individually integrated over all positions. The ordering of interest, in which three neutrons are first created, and then later destroyed, will be picked out by the choice of $E$, as discussed shortly. Second, we stress that $\mathcal N$ and $\overline {\mathcal N}$ are anticommuting Grassmann fields and that time-ordering is also subject to the usual Grassmann definition, e.g.~$\text{T}\{ \mathcal N(x^0) \mathcal N(y^0) \} = - \mathcal N(y^0) \mathcal N(x^0) $ for $x^0 < y^0$. The anticommuting property plays an important role in the formalism, as already noted above in the result for $\textbf G$, \Cref{eq:Glab}.

In this work we assume that $C_L(P)$ can be represented as an all-orders Feynman diagrammatic expansion in a relativistic effective field theory (EFT) in which the degrees of freedom are neutrons, protons and pions.
We make no restrictions on the form or strength of the interaction vertices, other than requiring that they respect Lorentz invariance, baryon number, and electric charge conservation.
We thus expect that the resulting formalism applies nonperturbatively.
Close to the three-neutron threshold, the pions could be integrated out, leading to a pionless EFT for nucleons,
but we do not need to assume that we are in this regime.
What we do assume is that the total CMF energy, $E_3^* = \sqrt{P^2} = \sqrt{E^2 - \boldsymbol P^2}$,
lies in a regime such that the only on-shell states are those involving three neutrons.
In particular, for isosymmetric QCD we require that
\begin{equation}
\sqrt{4 m_N^2 - m_\pi^2} + m_N < E_3^* < 3 m_N + m_\pi \,.
\label{eq:limitsonE3}
\end{equation}
The upper limit is required to avoid $3N+\pi$ on-shell states, while the lower limit avoids the left-hand cut in two-to-two subprocesses due to single-pion exchange. The latter restriction is necessary because, as explained in the following, the non-analyticity in $\textbf K_2$ from the left-hand cut leads to power-law finite-volume dependence.
We stress that, although proton degrees of freedom are present in the EFT, they cannot appear in on-shell
intermediate states in the three-neutron correlator for the kinematic regime of \Cref{eq:limitsonE3}.
Instead, they appear in virtual intermediate states—along with pions that lead, for example, to the dressing
of the neutron propagator—as well as in Bethe-Salpeter kernels.

Our aim in the following is to keep track of all power-like dependence of $C_L(P)$ on $L$ (typically of the form of powers of $1/L$), while neglecting dependence falling faster than any power of $1/L$. With a slight abuse of nomenclature, we will refer to the latter as ``exponentially suppressed''. Indeed, this category includes exponentially-suppressed scaling of the form $e^{- m_\pi L}$, where $m_\pi$ is the mass of the lightest degree of freedom in the system, e.g. the pion in QCD, and generally not the mass of the spin-$1/2$ particle, which we denote $m$ in the following. For large enough $L$, such exponentially-suppressed terms are numerically smaller than power-law effects.

The key difference between contributions to the finite- and infinite-volume correlators is that the former involve sums over finite-volume momenta, while the latter involve integrals. Local vertices are unchanged. As discussed in ref.~\cite{\HSQCa}, one can use the Poisson summation formula to argue that the difference between finite-volume sums and infinite-volume integrals is exponentially suppressed unless the summand/integrand is singular. This can happen either because there is an on-shell intermediate state, or because of a  non-analyticity in the vertex functions such as the above-mentioned left-hand cut. Since in the following derivation we include the effects only of three-neutron intermediate states, we are led to the same requirements on $E_3^*$ as given in \Cref{eq:limitsonE3}. We stress that the localization of $\cO$ in time plays an important role here, for it implies that states consisting, say, of two neutrons and an antineutron, cannot propagate forward in time for an arbitrarily long extent, and thus cannot lead to on-shell singularities.

The diagrammatic expansion involves the operators $\mathcal O$ and $\mathcal O^\dagger$ as well as the above-mentioned vertices. Simple examples are shown in \Cref{fig:deri}. The neutron propagators in these diagrams are given by
\begin{equation}
\Delta_{L, \alpha \beta}(p) =
\int_L d^4x \, e^{i p \cdot x } \, \langle \text{T} \mathcal N_{\alpha}(x) \overline{\mathcal N}_{\beta}(0) \rangle_L \,.
\end{equation}
They are thus fully dressed, and include loop diagrams that are not shown explicitly. The subscript $L$ on the expectation value refers to the $L$-dependence that arises from the spatial periodicity, which implies that $\bm p$ must be drawn from the finite-volume set, and that the spatial parts of loop momenta must be summed. However, in the vicinity of the single-particle pole at $p^2=m^2$, all loop contributions are far off shell, so that $L$-dependence is exponentially suppressed, and we can write \cite{Luscher:1986n1}
\begin{equation}
\label{eq:sep_single_pole}
\Delta_{L, \alpha \beta}(p) = i \frac{ ( \slashed p + m)_{\alpha \beta} }{p^2 - m^2 + i \epsilon} + R_{L, \alpha \beta}(p) \,.
\end{equation}
Here the first term is $L$-independent, while the second term is analytic for $p^0$ below the threshold to create a multi-particle state (a nucleon+pion state in the case of isospin-symmetric QCD). We have assumed a normalization of the fields $ \mathcal N$ and $\overline{\mathcal N}$ so as to give unit residue at the pole of the dressed propagator, leading to a simple factor of $\slashed p+m$ in the numerator.

One can apply a final simplification to the numerator of the $L$-independent part of the propagator by setting $p^0$ to its on-shell value: $p^0 = \omega_p = \sqrt{\boldsymbol p^2+m^2}$. This is justified since the difference cancels the pole, leading to an additional analytic contribution that can be absorbed by a redefinition of $ R_{L, \alpha \beta}(p) $. This step is useful as the on-shell version of the numerator is equal to a sum of spinors,
\begin{equation}
\label{eq:dirac_on_shell}
( \slashed p + m)_{\alpha \beta} \bigg \vert_{p^0 = \omega_p} = \sum_r u^r_{\alpha} (\boldsymbol p) \overline u^r_{\beta}(\boldsymbol p) \,.
\end{equation}
As noted in the previous section, and discussed further below, it is for this reason that the finite-volume factors $\textbf F$ and $\textbf G$ are proportional to Kronecker deltas in spin space in the {\em lab-axis frame}.

Like the momentum sums in the fully-dressed propagator, loop sums in generic Feynman diagrams for $C_L$ contribute only exponentially suppressed $L$-dependence, as long as the loops do not allow three-neutron cuts. Such cuts lead to poles in the summand, and induce power-like volume effects that must be included in (and indeed are the focus of) the derivation. Heuristically, the intuition is that these physical, on-shell intermediate states can travel arbitrarily far and thus maximally feel the boundary conditions.

From these considerations, it follows that a skeleton expansion in which all three-neutron states are explicitly displayed will capture the power-like $L$ dependence in $C_L(P)$. Such an expansion is built from all diagrams with operator-dependent vertex functions on the far left and far right, with three neutrons propagating between them, interacting via Bethe-Salpeter kernels. Examples are shown in \Cref{fig:deri}. The two-to-two kernels shown are defined to be two-particle irreducible in the s channel, so that there are no hidden three-particle states. There can also be three-to-three Bethe-Salpeter kernels, which are three-particle irreducible in the s channel. We discuss these kernels further in the remainder of the derivation.

\subsubsection{Contributing diagrams\label{sec:contributingdiagrams}}

\begin{figure}
\begin{center}
\includegraphics[width=0.8\textwidth]{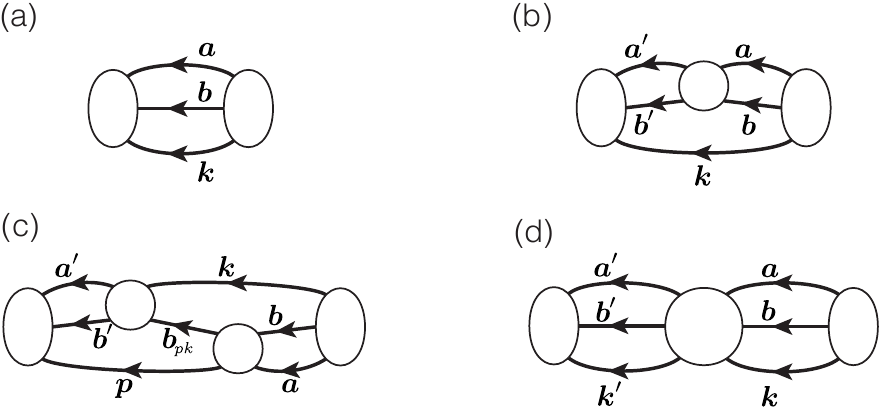}
\end{center}
\caption{Diagrams relevant for the derivation of the quantization condition for three neutrons in a finite volume. Lines with arrows indicate fully-dressed neutron propagators. (We envision time flowing from right to left to match the ordering of incoming and outgoing states in the expressions of the main text.) The tall ellipses indicate the $\sigma$ functions arising from the interpolating operators while the circles indicate two- and three-particle Bethe-Salpeter kernels.}\label{fig:deri}
\end{figure}

Continuing to follow the basic structure of the argument of ref.~\cite{\isospin}, we now analyze the three diagrams shown in \Cref{fig:deri} in turn, beginning with \ref{fig:deri}(a). Denoting the contribution of this to $C_L$ as $C_L^{[\ref{fig:deri}(a)]}$, we have
\begin{equation}
\label{eq:CL1a_def}
C_L^{[\ref{fig:deri}(a)]} = \frac{1}{6} \int_{a,\,L} \, \int_{k,\,L} \,
\sigma_{\alpha_1 \alpha_2 \alpha_3} (a,k) \Delta_{L, \alpha_1 \beta_1}(a)
\Delta_{L, \alpha_2 \beta_2}(b) \Delta_{L, \alpha_3 \beta_3}(k)
\sigma^\dagger_{\beta_1 \beta_2 \beta_3}(a,k) \,,
\end{equation}
where $b = P-k-a$, and we have introduced the ``endcap'' $ \sigma_{\alpha_1, \alpha_2 ,\alpha_3} (a,k) $, defined by
\begin{equation}
\label{eq:sigma_func_def}
\sigma_{\alpha_1 \alpha_2 \alpha_3} (a,k) = \sum_{{\sf p} \in \mathcal P} \text{sig}({\sf p}) \times \overline u^{r_1}_{\alpha_1}(\boldsymbol a) \overline u^{r_2}_{\alpha_2}(\boldsymbol b) \overline u^{r_3}_{\alpha_3}(\boldsymbol k) \times {\sf p} \! \left [ f_{r_1r_2r_3}(a,b,k) \right ] \,.
\end{equation}
Here $\mathcal P$ is the set of six permutations acting on the momentum and spin indices. For example if we define ${\sf p}_{1 \leftrightarrow 2}$ as the permutation exchanging both the indices $r_1$ and $r_2$ and the corresponding momenta $a$ and $b$, then
\begin{equation}
{\sf p}_{1 \leftrightarrow 2} \! \left [ f_{r_1r_2r_3}(a,b,k) \right ] = f_{r_2r_1r_3}(b,a,k) \,.
\end{equation}
The factor $\text{sig}({\sf p})$ is $1$ for a cyclic permutation and $-1$ otherwise. As a result, only the antisymmetric part of $ f_{r_1r_2r_3}(a,b,k) $ contributes in \Cref{eq:sigma_func_def}.

To analyze \Cref{eq:CL1a_def}, one next uses the identity of \Cref{eq:sep_single_pole} to evaluate the $k^0$ and $a^0$ integrals within $\int_{a,\,L}$ and $ \int_{k,\,L}$. This yields
\begin{multline}
\label{eq:CL1a_v2}
C_L^{[\ref{fig:deri}(a)]}(P) = \widetilde C_{\infty}^{[\ref{fig:deri}(a)]}(P)
\\
+ \frac{1}{6} \frac{1}{L^6} \sum_{\boldsymbol a, \boldsymbol k}
\sigma_{\alpha_1 \alpha_2 \alpha_3} (\bm a, \bm k)
\frac{i ( \slashed a + m)_{\alpha_1 \beta_1} \, ( \slashed b + m)_{\alpha_2 \beta_2} \,
( \slashed k + m)_{\alpha_3 \beta_3} }{2 \omega_a 2 \omega_k (b^2-m^2)}
\sigma^\dagger_{\beta_1 \beta_2 \beta_3}(\bm a,\bm k)
\,,
\end{multline}
where $\widetilde C_{\infty}^{[\ref{fig:deri}(a)]}(P) $ is a quantity with negligible (exponentially suppressed) $L$ dependence. (The tilde is used as we will require a redefinition to reach our final quantity, $ C_{\infty}^{[\ref{fig:deri}(a)]}(P) $.)

To reach \Cref{eq:CL1a_v2} we have used the result that all contributions containing at least one factor of $R_{L, \alpha \beta}(p)$ lead to exponentially suppressed volume dependence. In the term proportional to $1/[(a^2-m^2)(b^2-m^2)(k^2-m^2)]$ we have evaluated the $a^0$ and $k^0$ integrals by closing the contours in the complex plane, encircling poles at $a^0 = \omega_a$ and $k^0 = \omega_k$. This also results in the numerator factors $( \slashed a + m)$ and $( \slashed k + m)$ being evaluated on-shell, allowing the subsequent use of the result \Cref{eq:dirac_on_shell}, as explained below. We also note that the four-momenta in the arguments of $\sigma$ and $\sigma^\dagger$ are set on shell, and, with some abuse of notation, we denote this by changing the arguments to three-vectors. All other poles that contribute to the $a^0$ and $k^0$ integrals can be seen to lead to intermediate states that cannot go on shell for our range of $E_3^*$, and thus give nonsingular summands, for which the sums can be replaced by integrals, so that the contributions can be absorbed in $\widetilde C_{\infty}^{[\ref{fig:deri}(a)]}(P)$.

The remaining momentum $b^\mu = (E - \omega_k - \omega_a , \boldsymbol P - \boldsymbol k - \boldsymbol a)$ is generally not on the mass shell, but the summand can be expanded about the on-shell point, i.e.~about $E = \omega_k + \omega_a + \omega_b$,. This expansion, together with a few additional manipulations explained below, yields
\begin{multline}
\label{eq:CL1a_v3}
C_L^{[\ref{fig:deri}(a)]}(P) = C_{\infty}^{[\ref{fig:deri}(a)]}(P)
+ \frac{1}{6} \frac{1}{L^3} \sum_{ \boldsymbol k} H(\boldsymbol k) \bigg [ \frac{1}{L^3} \sum_{\boldsymbol a} -
{\rm p.v.} \int_{\boldsymbol a} \bigg ]
\\
\times
\sigma^{\sf lab}_{r_1 r_2 r_3} (\bm a, \bm k)
\frac{i \delta_{r_1r_1'} \delta_{r_2r_2'} \delta_{r_3r_3'} }
{2 \omega_a 2 \omega_b 2 \omega_k (E - \omega_a - \omega_b - \omega_k) }
\sigma^{{\sf lab},\dagger}_{r'_1 r'_2 r'_3}(\bm a,\bm k)
\,,
\end{multline}
where
\begin{align}
\sigma^{\sf lab}_{r_1 r_2 r_3} (\bm a,\bm k) &=
\sigma_{\alpha_1 \alpha_2 \alpha_3} (\bm a,\bm k) \,
{u}_{\alpha_1}^{r_1}(\bm a) {u}_{\alpha_2}^{r_2}(\bm b) {u}_{\alpha_3}^{r_3}(\bm k) \,,
\\
&=
(2m)^3 \sum_{{\sf p} \in \mathcal P} \text{sig}({\sf p}) {\sf p} \! \left [ f_{r_1r_2r_3}(\bm a,b, \bm k) \right ] \,.
\end{align}
In the second line, the middle argument of $f$ remains a four-vector $b$, since this momentum is not yet on shell. The superscript ${\sf lab}$ emphasizes that the spin components are defined with respect to the lab-frame axis. The additional steps to reach \Cref{eq:CL1a_v3} include the subtraction of the integral, which is then added back in through a redefinition of $C_{\infty}^{[\ref{fig:deri}(a)]}(P)$, which is allowed since the p.v.\ integral yields a smooth function of $\bm k$, allowing the sum over $\bm k$ to be replaced by an integral. A similar redefinition allows us to introduce the cutoff function $H(\boldsymbol k)$, (whose form is discussed following \Cref{eq:Kell} above). These steps are explained in detail in the original derivation of the quantization condition~\cite{\HSQCa}. We can now use the expansion of the summand about the on-shell point to replace $\slashed b+m$ and other factors, with their on-shell values, for the difference cancels the pole and leads to a sum-integral difference that is exponentially suppressed. We can then apply \Cref{eq:dirac_on_shell} to the Dirac factors to write each as a product of spinors, leading to the expression shown for $\sigma^{\sf lab}$.

At this stage we could directly use the sum-integral identity given in eq.~(A1) of ref.~\cite{\HSQCa} to rewrite $C_L^{[\ref{fig:deri}(a)]}(P) $ in terms of the geometrical function $\textbf F^{\sf lab}$ given in \Cref{eq:Fspin},
\begin{equation}
\label{eq:diagram_a_lab_final}
C_L^{[\ref{fig:deri}(a)]} = C_\infty^{[\ref{fig:deri}(a)]} - \frac{1}{3}
\boldsymbol{\sigma}^{\sf lab} \textbf F^{\sf lab} \boldsymbol{\sigma}^{{\sf lab}, \dagger} \,.
\end{equation}
The overall $1/3$ arises because $\textbf F^{\sf lab}$ contains a factor of $1/2$, and the product leads to the $1/6$ in \Cref{eq:CL1a_v3}. The bold-faced $\boldsymbol{\sigma}^{\sf lab}$ indicates that a factor of $i$ has been absorbed into $\sigma$, that this quantity has been set on shell, and projected onto spherical harmonics using \Cref{eq:PWproj}. $\boldsymbol{\sigma}^{\sf lab}$ also includes the lab-frame axis spin indices, which, as shown in \Cref{eq:CL1a_v3}, are unchanged by the three-particle cut.%
\footnote{%
Note that we use a different labelling of the spin indices here ($r_i$) compared to those used in the definition of $\textbf F^{\sf lab}$ ($\bm m_s$), but this is just for notational convenience---the indices are the same.
}

However, this form is not used in our analysis, which is carried out in the dimer-axis frame. Instead, we must first rotate to the dimer-axis frame using the transformation given in \Cref{eq:basisspinrotation,eq:Dka}; and then apply the sum-integral identity discussed above, which involves projecting the resulting endcaps onto spherical harmonics using \Cref{eq:PWproj}, and then setting them on shell. These steps lead to
\begin{equation}
\label{eq:diagram_a_final}
C_L^{[\ref{fig:deri}(a)]} = C_\infty^{[\ref{fig:deri}(a)]}
- \frac{1}{3} \boldsymbol{\sigma} \textbf F \boldsymbol{\sigma}^{\dagger } \,,
\end{equation}
i.e. a result with the same form as the lab-axis frame version \Cref{eq:diagram_a_lab_final}. We stress that this identity of form hides a nontrivial difference. In particular, since the transformation connecting the lab- and dimer-axis frames depends on the momenta, it does not commute with the projection onto spherical harmonics. Thus, one cannot directly relate \Cref{eq:diagram_a_lab_final,eq:diagram_a_final}, but must instead must go back to the starting point of \Cref{eq:CL1a_v3}.

We now explain these steps in more detail. The basis rotation involves one subtlety, namely that the momentum $b$ is off shell, while the rotation described in \Cref{eq:basisspinrotation,eq:Dka} assumes that all momenta are on shell. Thus, we must extend the definition of the rotation to the off-shell case, and this we do by setting $\boldsymbol \beta_b= \bm b/\omega_b$ in the definition of $R_b$ needed in \Cref{eq:Dka}. This is the same extension already used in \Cref{sec:Gsub} when describing the form of $G$.

The second step is to use the sum-minus-integral identity from ref.~\cite{\HSQCa}, leading to endcaps that are projected, using \Cref{eq:PWproj}, onto definite orbital angular momentum $\ell m$ in the dimer CMF, and then set fully on shell. Since $\boldsymbol k$ is only sampled at discrete finite-volume values, it can also be viewed as an index, such that $\sigma$ is promoted to a vector in the combined index space of $r_1, r_2, r_3, \ell, m, \boldsymbol k$, where $\{r_i\}$ are spin indices in the dimer-axis frame, which were denoted $\{m_s(\bm k), m_s(\bm a^*), m_s(\bm b^*)\}$ earlier. We abbreviate the quantity labelled by all these indices as $\boldsymbol{\sigma}$ (with the symbol in bold), which also includes an additional factor of $i$. This leads to \Cref{eq:diagram_a_final}, in which the second term on the right-hand side is a matrix product built from the vector $\boldsymbol\sigma$ and its conjugate, together with the matrix $\textbf F $ defined in \Cref{eq:Flab}. This result has the same form as in the case of three pions of arbitrary isospin~\cite{\isospin}, except that here the flavor indices are replaced with (dimer-axis) spin indices.

We now turn to \Cref{fig:deri}(b). In this diagram a Bethe-Salpeter kernel, denoted $\mathcal B$, is inserted to scatter two of the three neutrons propagating across the diagram. The functional form is given by
\begin{multline}
\label{eq:diagB}
C^{[\ref{fig:deri}(b)]}_L(P) = \frac{1}{4} \frac{1}{L^9} \sum_{\boldsymbol k, \boldsymbol a, \boldsymbol a'}
\int \frac{d a^0}{2 \pi} \int \frac{d a'^0}{2 \pi} \int \frac{d k^0}{2 \pi}
\sigma_{\alpha_1 \alpha_2 \alpha_3}(a',k) \, \Delta_{\alpha_1 \alpha_1'}(a')
\Delta_{\alpha_2 \alpha_2'}(b') \delta_{\alpha_3 \alpha_3'}
\\
\times i \mathcal B_{\alpha'_1 \alpha'_2, \beta_1' \beta_2'}(a', b'; a, b) \,
\Delta_{\alpha_3', \beta_3'}(k) \, \Delta_{\beta_1' \beta_1}(a) \Delta_{\beta_2' \beta_2}(b) \,
\delta_{\beta_3' \beta_3}\, \sigma^\dagger_{\beta_1 \beta_2 \beta_3}(a,k) \,.
\end{multline}
Now, in contrast to the analysis of \Cref{fig:deri}(a), here we only aim to identify the maximally singular contribution to the diagram. The less singular parts will be absorbed into re-definitions of factors such as $\boldsymbol{\sigma}$ and $\boldsymbol{\sigma}^\dagger$ appearing in \Cref{eq:diagram_a_final}.

Because \Cref{fig:deri}(b) admits two distinct cuts across three-neutron states, the maximally singular contribution contains two poles. To isolate these we perform the $k_0$, $a'^0$, and $a^0$ contour integrals and expand the partially on-shell propagators about the physical energy points $E = \omega_{a'}+\omega_{b'}+\omega_k$ and $E = \omega_{a}+\omega_{b}+\omega_k$ to reach
\begin{multline}
C_L^{[\ref{fig:deri}(b)]}(P) = \frac{1}{4} \frac{1}{L^3}
\sum_{ \boldsymbol k} \bigg [ \frac{1}{L^3} \sum_{\boldsymbol a'} - \int_{\boldsymbol a'} \bigg ]
\bigg [ \frac{1}{L^3} \sum_{\boldsymbol a} - \int_{\boldsymbol a} \bigg ]
\\
\times
\sigma^{\sf lab}_{r_1 r_2 r_3} (\bm a', \bm k)
\frac{i \delta_{r_1r_1'} \delta_{r_2r_2'} \delta_{r_3r_3'} H(\boldsymbol k) }
{2 \omega_{a'} 2 \omega_{b'} 2 \omega_k (E - \omega_{a'} - \omega_{b'} - \omega_k) }
i [2 \omega_k] \mathcal B^{\sf lab}_{r_1' r_2', s_1' s_2'}(\bm a', b'; \bm a, b) \delta_{r_3' s'_3}
\\
\times \frac{i \delta_{s_1' s_1} \delta_{s_2' s_2} \delta_{s_3' s_3} H(\boldsymbol k) }
{2 \omega_a 2 \omega_b 2 \omega_k (E - \omega_a - \omega_b - \omega_k) }
\sigma^{{\sf lab}, \dagger}_{s_1, s_2 ,s_3}(\bm a,\bm k) + \cdots
\,,
\end{multline}
where the ellipsis denotes less singular terms that we do not track explicitly. Here we have already mimicked the steps performed for \Cref{fig:deri}(a) above including (i) replacing the sums over $\boldsymbol{a}$ and $\boldsymbol{a}'$ with integrals and sum-integral differences (and discarding the integrals as these lead to less singular terms) (ii) inserting the identity $1 = H(\boldsymbol k) + [1 - H(\boldsymbol k)]$ and again dropping the second contribution as it is less singular (iii) expanding all Dirac structures about the on-shell point to recover outer products of spinors that are then contracted with the endcaps and with the Bethe-Salpeter kernel.

The final steps, again in a direct imitation of the analysis for \Cref{fig:deri}(a), are to expand the endcap factors and Bethe-Salpeter kernel about their on-shell points, to perform a momentum-dependent change of basis to the dimer-axis frame, and then to decompose all quantities in angular momentum in the dimer-axis frame. The result takes a very compact form
\begin{equation}
C_L^{[\ref{fig:deri}(b)]} = - \boldsymbol{\sigma} \, \textbf F \, \textbf B_2 \, \textbf F \, \boldsymbol{\sigma}^{\dagger} + \cdots \,,
\end{equation}
where $\textbf B_2$ is analogous to $\textbf K_2$, defined in \Cref{eq:boldK2_spin_def} above, but based on the Bethe-Salpeter kernel rather than the full K matrix.

\bigskip

We finally reach \Cref{fig:deri}(c), which we also write out in detail to fully explain the pattern of the derivation that we are identifying. The diagram can be expressed as
\begin{equation}
\label{eq:diagC}
\begin{split}
C^{[\ref{fig:deri}(c)]}_L(P)
= (-1)
\frac{1}{4} \frac{1}{L^{12}}
\sum_{\boldsymbol k, \boldsymbol p, \boldsymbol a, \boldsymbol a'}
\int \frac{d a^0}{2 \pi} & \int \frac{d a'^0}{2 \pi} \int \frac{d k^0}{2 \pi} \int \frac{d p^0}{2 \pi}
\\[5pt] \times \bigg [ & \sigma_{\alpha_1, \alpha_2, \alpha_3}(a',p) \,
\Delta_{\alpha_1 \alpha_1'}(a') \Delta_{\alpha_2 \alpha_2'}(b') \delta_{\alpha_3 \alpha_3'}
\\[5pt] & \times
i \mathcal B_{\alpha'_1 \alpha'_2, \beta_1' \beta_2'}(a', b'; k, b_{pk}) \, \Delta_{\alpha_3' \beta_3'}(p) \,
\\[5pt] & \times
\Delta_{\beta_2' \gamma_2'}(b_{pk}) \delta_{\beta_1' \gamma_3'} \delta_{\beta_3' \gamma_1'}\, \\[5pt]
& \times i \mathcal B_{\gamma'_1 \gamma'_2 \sigma_1' \sigma_2'}(p, b_{pk}; a, b) \, \Delta_{\gamma_3' \sigma_3'}(k) \\[5pt] & \times
\Delta_{\sigma_1' \sigma_1}(a) \Delta_{\sigma_2' \sigma_2}(b) \, \delta_{\sigma_3' \sigma_3}\, \sigma^\dagger_{\sigma_1 \sigma_2 \sigma_3}(a,k) \bigg ]\,,
\end{split}
\end{equation}
where the initial $(-1)$ arises from the anti-commutation of Grassmann fields in the correlation function defining the diagram.

Each line of the lengthy expression in square brackets corresponds to a specific segment of the diagram. In particular, the first and last lines correspond to the two endcap factors, each with the two propagators that are contracted between an endcap and its adjacent Bethe-Salpeter kernel. The second and fourth lines then correspond to the two kernels themselves, each accompanied by the spectator propagator, i.e.~the propagator that is not scattered by that kernel. Finally, the middle line corresponds to the distinctive feature of \Cref{fig:deri}(c), namely the change in the particle that is spectating. This switch leads to the $b_{pk}$-dependent propagator as well as the set of Kronecker deltas, $\delta_{\beta_1' \gamma_3'} \delta_{\beta_3' \gamma_1'}$, that encode the interchange of indices with $1$ and $3$ subscripts.

Again identifying the most singular term in the expression, and performing steps discussed in detail in ref.~\cite{\HSQCa}, we reach
\begin{equation}
\begin{split}
C_L^{[\ref{fig:deri}(c)]}(P) &
=
\frac{1}{4} \frac{1}{L^6} \sum_{ \boldsymbol p, \boldsymbol k}
\bigg [ \frac{1}{L^3} \sum_{\boldsymbol a'} - \int_{\boldsymbol a'} \bigg ]
\bigg [ \frac{1}{L^3} \sum_{\boldsymbol a} - \int_{\boldsymbol a} \bigg ]
\\
& \times
\sigma^{\sf lab}_{r_1 r_2 r_3} (\bm a',\bm p)
\frac{i \delta_{r_1r_1'} \delta_{r_2r_2'} \delta_{r_3r_3'} H(\boldsymbol p)}
{2 \omega_{a'} 2 \omega_{b'} 2 \omega_p (E - \omega_{a'} - \omega_{b'} - \omega_p) }
\\
& \times i [2 \omega_p] \mathcal B^{\sf lab}_{r_1' r_2', s_1' s_2'}(\bm a', b'; \bm k, b_{pk}) \delta_{r_3' s'_3}
\\
& \times (-1) \delta_{s'_2 t'_2} \delta_{s'_1 t'_3} \delta_{s'_3 t'_1}
\frac{i}{4 \omega_p \omega_k } \frac{H(\boldsymbol{p}) H(\boldsymbol{k})}{b_{pk}^2-m^2}
\\
& \times i[ 2 \omega_k] \mathcal B^{\sf lab}_{t_1' t_2', z_1' z_2'}(\bm p, b_{pk} ; \bm a, b) \delta_{t_3' z_3'}
\\
& \times \frac{i \delta_{z_1' z_1} \delta_{z_2' z_2} \delta_{z_3' z_3} H(\boldsymbol k) }
{2 \omega_a 2 \omega_b 2 \omega_k (E - \omega_a - \omega_b - \omega_k) }
\sigma^{{\sf lab}, \dagger}_{z_1 z_2 z_3}(\bm a, \bm k)
\\
& + \cdots
\,.
\end{split}
\end{equation}
The key observation here is that all structures appearing in this long equation have already been used above except for the line beginning with $\times (-1) \delta_{s'_2 t'_2} \delta_{s'_1 t'_3} \delta_{s'_3 t'_1}\cdots$. Again this is the line that encodes the switch in which pair is scattering.

We once again repeat the now usual steps of expanding about the on-shell point, performing the momentum-dependent change of basis to the dimer-axis frame and decomposing all quantities in spherical harmonics. Because the two-particle momenta differ on either side of the switch factor, the change of basis matrices do not cancel. This leads to the nontrivial form of $\textbf G$ given in \Cref{eq:Gspin}. In our notation, the final result for the most singular-part of \Cref{fig:deri}(c) takes a very compact form:
\begin{equation}
C_L^{[\ref{fig:deri}(c)]} = - \boldsymbol{\sigma} \, \textbf F \, \textbf B_2 \, \textbf G \, \textbf B_2 \, \textbf F \, \boldsymbol{\sigma}^{\dagger} + \cdots
\end{equation}
\bigskip

This completes our analysis of the three key diagrams of \Cref{fig:deri}. In the following section we illustrate how these lead to a generalized quantization condition for finite-volume three-nucleon systems.

\subsubsection{All orders generalization\label{sec:allorders}}

To reach our final form for $C_L(P)$, the correlation function introduced in \Cref{eq:CLdef}, we need to analyze all contributions to the skeleton expansion. We stress that in the following we are explaining the logic and structure of the full derivation, but many details are left out. For these we refer to the derivation for identical particles given in refs.~\cite{\HSQCa,\HSQCb}. One feature that changes compared to that derivation is the need for amplitudes involving three identical fermions to be antisymmetric under particle exchange, rather than symmetric. The impact of this change is discussed in \Cref{app:deriv}.

An important building block entering the expansion that we have not considered so far is the three-particle Bethe-Salpeter kernel. An example of a diagram containing this kernel is shown in \Cref{fig:deri}(d). When represented as a function of momenta and spin indices, the kernel is denoted by $\mathcal B_3$ in analogy to the two-particle Bethe-Salpeter kernel, $\mathcal B$ [appearing, for example, in \Cref{eq:diagB}]. $\mathcal B_3$ is defined as the sum over the set of all Feynman diagrams with three incoming and three outgoing neutrons that do not contain any three-neutron cuts intersecting the flow of total energy and momentum ($s$-channel cuts).%
\footnote{%
Unlike for the case of pions, there are no diagrams with s-channel cuts containing a single neutron, due to baryon-number conservation. Thus, an additional complication in the construction of $\cB_3$ that was addressed for pions and similar particles in ref.~\cite{\HSQCa} is avoided here.
}

$\mathcal B_3$ depends on the same kinematical and spin variables as $\mathcal K_{\text{df},3}$, allowing the description in \Cref{sec:Kdfsub} to carry over. If the outgoing spin indices and the outgoing momenta of $\mathcal B_3$ are fixed, then the incoming degrees of freedom exactly match those of $\sigma$, the leftmost endcap function used extensively in the previous section. From this it follows that the steps leading to the on-shell projection of $\sigma$ and the rotation to the dimer-frame axis can also be followed for $\mathcal B_3$. If these steps are applied to both the incoming and outgoing kinematics, then $\mathcal B_3$ can be represented as a matrix in the same space as $\textbf K_2$, $\textbf F$, and $\textbf G$. After absorbing a factor of $i$ this matrix is denoted by $\textbf B_3$.

Returning to the main analysis, we follow the same logic as in the original RFT derivation for identical spin-zero particles \cite{\LtoK} by noting that every diagram generates a maximally singular piece as well as contributions with fewer singularities. The latter arise when the loop momenta associated with poles are integrated using the principal value prescription rather than summed, and from finite terms that result when expanding the coefficients of poles about their on-shell values. Such nonmaximally singular terms can be incorporated through redefinitions of infinite-volume quantities $\boldsymbol \sigma$, $\textbf B_2$, and $\textbf B_3$. For example, \Cref{fig:deri}(a) has a single pole in its maximally singular contribution whereas \Cref{fig:deri}(b) (which has two poles in its maximally singular term) can also generate single-pole contributions when either $\boldsymbol a$ or else $\boldsymbol a'$ (but not both) is integrated. The single-pole parts of \Cref{fig:deri}(a) and (b) are ultimately combined through redefinitions of $\boldsymbol \sigma$ and $\boldsymbol \sigma^\dagger$, in a step that collects an infinite set of single-pole terms into a single contribution to $C_L(P)$.

We now identify all contributions to $C_L(P)$, beginning with the rule for identifying the maximally singular piece in each diagram.%
\footnote{%
We emphasize that the presentation here follows a different strategy for reviewing the derivation of the RFT quantization condition, as compared, e.g., to refs.~\cite{\LtoK,Hansen:2020zhy}. This difference is not related to the consideration of particles with spin; indeed an analogous approach could also have been applied in earlier works. We think that the present approach allows a straightforward understanding of the essential features of the final results, although it hides many details explained in ref.~\cite{\HSQCa}.
}
To explain the rule it is convenient to first introduce the notion of a \emph{two-to-two sub-process sector}, defined as any section of skeleton-expansion diagram that appears between insertions of $\boldsymbol \sigma$, $\textbf B_3$ or $\boldsymbol \sigma^\dagger$ (and does not contain any such factors). For example, the three diagrams analyzed so far each contain exactly one two-to-two subprocess sector as no $\textbf B_3$ factors are included. Within each such sector, one can identify the number of times that the pair of scattering nucleons changes (the number of switches, $N_{\sf{switch}}$) or, similarly, the number of elastic two-to-two sections, $N_{{\sf elast}} = N_{\sf{switch}}+1$, in which pairwise scattering occurs for a given nucleon pair. We also adopt the convention that $N_{{\sf elast}} = 0$ if no pairwise scattering occurs in a particular two-to-two subprocess sector, as in \Cref{fig:deri}(a).

The maximally singular contribution of any subprocess sector, denoted by $\textbf S^{( \boldsymbol n)}_L(P)_{\sf max}$, then takes a simple form
\begin{equation}
\label{eq:SLpart1}
\textbf S^{( \boldsymbol n)}_L(P)_{\sf max} =
\begin{cases}
\textbf{F}/3 & N_{{\sf elast}} = 0 \,,
\\
\! \bigg (\! \prod_{k=1}^{N_{{\sf elast}}} \!
\Big ( \big [ \textbf F \textbf B_2 \big ]^{\boldsymbol n(k)} \textbf G {\textbf B}_2 \Big ) \bigg ) \textbf B_2^{-1} \textbf G^{-1}
\textbf F & N_{{\sf elast}} > 0 \,,
\end{cases} \quad \text{(maximally\ singular)},
\end{equation}
where $\boldsymbol n$ is a vector of length $N_{{\sf elast.}}$, populated with positive integers counting the number of $\textbf B_2$ factors in each elastic two-to-two section, and $\boldsymbol n(k)$ is its ${k}^\text{th}$ component. Here it is understood that the $k=1$ factor is on the far left and the $k = N_{{\sf elast.}}$ on the far right. Note that $\textbf G$ appears $N_{{\sf elast.}} - 1 = N_{\sf{switch}}$ times.

In \Cref{eq:SLpart1} we have used the subscript ${\sf max}$ to indicate that the right-hand side is only part of the contribution. This is because we intend $ \textbf S^{(\boldsymbol n)}_L(P)$, with no ${\sf max}$ subscript, to contain all contributions with a set of poles described by a given $\boldsymbol n$. We are missing the non-maximally-singular terms that arise when some coordinates are integrated, as well as those associated with expanding the coefficients of poles. As is shown in ref.~\cite{\LtoK}, including the former has the effect of promoting the Bethe-Salpeter kernel, $\textbf B_2$, to the K-matrix, $\textbf K_2$, defined in \Cref{sec:K2sub} and in particular in \Cref{eq:boldK2_spin_def} above. These can thus be incorporated by making the simple adjustment
\begin{equation}
\textbf S^{(\boldsymbol n)}_L(P) =
\begin{cases}
\textbf{F}/3 & N_{{\sf elast}} = 0 \,, \\
\bigg ( \prod_{k=1}^{N_{{\sf elast}}}
\Big ( \big [ \textbf F \textbf K_2 \big ]^{\boldsymbol n(k)} \textbf G {\textbf K}_2 \Big ) \bigg )\textbf{ K}_2^{-1} \textbf G^{-1}
\textbf F & N_{{\sf elast}} > 0 \,,
\end{cases} \qquad \text{(all\ terms)}.
\end{equation}

The next step is to sum over $N_{\sf elast.}$ values and over all positive integer components of the vectors $\boldsymbol n$. This gives a quantity denoted by $\textbf F_3$:
\begin{align}
\textbf F_3 & = \sum_{N_{\sf elast.}=\,0}^\infty \ \sum_{\boldsymbol n \in (\mathbb Z^+)^{N_{\sf elast.}}} \ \textbf S^{(\boldsymbol n)}_L(P) \,.
\end{align}
Evaluating the sum explicitly, one finds
\begin{equation}
\textbf F_3 = \frac{\textbf F}{3} + \textbf F \frac{1}{1 - \textbf M_{2,L} \textbf G} \textbf M_{2,L} \textbf F \,, \qquad \qquad \textbf M_{2,L} = \frac{1}{\textbf K_2^{-1} - \textbf F} \,.
\label{eq:M2Ldef}
\end{equation}
This matches the definition in refs.~\cite{\LtoK,Hansen:2020zhy}.

A more subtle issue is the impact of finite terms obtained when the coefficients of poles are expanded about their on-shell values. We have done such an expansion, for example, in the analysis of \Cref{fig:deri}(c), when we replaced the pole with the factor of $\textbf G$, and set adjacent quantities on shell. The terms that are dropped effectively sew the adjacent Bethe-Salpeter kernels together creating a contribution to the three-particle K matrix. It turns out that such contributions can be absorbed into changes in $\textbf B_3$, although this requires the use of a symmetrization procedure that is discussed in \Cref{app:deriv}. This allows one to proceed by ignoring this issue for now, and adding in such terms after discussing the role of $\mathbf B_3$.

To explain this role, we iterate the insertion of $\textbf F_3$---corresponding to a two-to-two subprocess sector---with the objects that break up these sectors, namely $\boldsymbol{\sigma}$, $\textbf B_3$, and $\boldsymbol{\sigma}^\dagger$. This leads to the following naive result for $C_L(P)$,
\begin{equation}
\label{eq:B3series}
C_L(P) = C_\infty(P) - \sum_{n=0}^\infty \boldsymbol{\sigma} \textbf F_3 \big [ \textbf B_{3} \textbf F_3 \big ]^n \boldsymbol{\sigma}^\dagger \qquad (\mathrm{incomplete}) \,,
\end{equation}
where we have stressed that this summation is still incomplete. To complete it we need to include the above-discussed finite contributions from expanding pole coefficients about the on-shell point, as well as contributions in which momentum arguments $\boldsymbol{\sigma} $, $ \textbf B_{3} $, and $\boldsymbol{\sigma}^\dagger$ are integrated. The combined effect is that one must make replacements analogous to the $\textbf B_2 \to \textbf K_2$ replacement discussed above. Here these are as follows: $\textbf B_3 \to \textbf K_{\text{df},3}$, $\boldsymbol \sigma \to \textbf A'_3$, $\boldsymbol \sigma^\dagger \to \textbf A_3$. The first of the new quantities, $\textbf K_{\text{df},3}$, is discussed in \Cref{sec:Kdfsub}, and the full definition is given in \Cref{sec:inteqs} below via the integral equations that relate this quantity to the physical scattering amplitude. Similarly, as was shown in ref.~\cite{Hansen:2021ofl}, $\textbf A'_3$ and $\textbf A_3$ satisfy integral equations that define them in terms of matrix elements of the operators entering $C_L(P)$. These latter two quantities play no role in the quantization condition and are not discussed further in this work.

Putting everything together we find that \Cref{eq:B3series} is replaced with an equally simple series,
\begin{equation}
C_L(P) = C_\infty(P) - \sum_{n=0}^\infty \textbf A'_3 \textbf F_3
\big [ \textbf K_{\text{df},3} \textbf F_3 \big ]^n \textbf A_3 \,,
\end{equation}
that evaluates to the final form for the finite-volume correlator
\begin{equation}
\label{eq:CLresult_3N}
C_L(P) = C_\infty(P) - \textbf A'_3 \textbf F_3 \frac{1}{1 - \textbf K_{\text{df},3} \textbf F_3} \textbf A_3 \,.
\end{equation}
This result is symbolically equivalent to those of previous works on different three-particle systems, differing only in the detailed meaning of $\textbf K_{\text{df},3} $, $\textbf F_3$ and the various matrices making up the definition of the latter, in particular $\textbf F$, $\textbf G$ and $\textbf K_2$.

\bigskip

Before concluding this section we consider an alternative finite-volume correlation function that is useful for deriving the relation between the two- and three-body K-matrices and the physical three-to-three scattering amplitude. This is the generalization to spin-$1/2$ particles of the quantity $ \mathcal M_{3,L}^{(u,u)}$ introduced in \Cref{eq:M3Ldef}. It is a finite-volume analog of a scattering amplitude, with three incoming and three outgoing nucleons. It can be obtained from the diagrams contributing to $C_L(P)$ by removing the endcaps and amputating external propagators. One additional technicality is required in the case where the outermost interaction, adjacent to the incoming or outgoing state, is a two-to-two rather than a three-to-three Bethe-Salpeter kernel. In this case the incoming momentum $\boldsymbol k$ and the outgoing momentum $\boldsymbol p$ are always assigned to the external nucleon not connected to the outermost $\textbf B_2$ factor. Thus it is an unsymmetrized quantity, on both incoming and outgoing sides, as indicated by the superscript $(u,u)$. It can be written using the same matrix indices as for the matrices in the quantization condition. Absorbing a factor of $i$, the resulting matrix is denoted $\textbf M_{3,L}^{(u,u)}(P)$.

A finite-volume decomposition directly analogous to that for $C_L(P)$ can then be performed for this quantity and leads to the following result:
\begin{align}
\label{eq:M3Ldef_3N}
\textbf M_{3,L}^{(u,u)}(P) & \equiv \textbf D^{(u,u)} + \textbf L^{(u)}_L \frac{1}{1 - \textbf K_{\df,3} \textbf F_3} \textbf K_{\df,3} \textbf R^{(u)}_L \,,
\end{align}
where
\begin{align}
\textbf D^{(u,u)} & \equiv \frac{1}{1 - \textbf M_{2,L} \textbf G } \textbf M_{2,L} \textbf G \textbf M_{2,L} \,, \\
\textbf L^{(u)}_L & \equiv \textbf F^{-1} \cdot \textbf F_3 \,, \\
\textbf R^{(u)}_L & \equiv \textbf F_3 \cdot \textbf F^{-1}\,.
\end{align}
Note that these results have similar form as those for identical scalars, eqs.~\eqref{eq:Ddef_scalar}, \eqref{eq:Ldef_scalar}, and \eqref{eq:Rdef_scalar} above, with the differences resulting from the factors of $i$, $2\omega$ and $L^3$ that have been absorbed into our boldface definitions.

\section{Summary of results and implementation\label{sec:summary}}

In this section we present the final results from the derivation of the previous section. We also comment briefly on the issues that will arise when implementing the formalism.

\subsection{Relating finite-volume energies to \texorpdfstring{$\Kdf$}{the divergence-free K-matrix}}

In direct analogy to previous derivations, see e.g.~refs.~\cite{Kim:2005gf,\LtoK}, the condition that $C_L(P)$ contains a pole in $E$ (at fixed $\boldsymbol P$ and $L$) implies the following finite-volume quantization condition
\begin{equation}
\det_{\boldsymbol k, \ell, m, \boldsymbol m^*_s}\big [ 1 - \textbf K_{\df,3} \textbf F_3 \big ] = 0 \,.
\end{equation}
Here the subscripts on the determinant indicate that it is taken in the space on which all matrices are defined, which is a Kronecker-product space of the spectator momentum $\boldsymbol k$ (discretized by the finite volume as $\boldsymbol k \in (2 \pi/L) \mathbb Z^3$), the orbital angular momentum of the two-particle pair, $\ell, m$, as well as the azimuthal spin components in the {dimer-axis frame}, $\boldsymbol m_s^*$. This result has the by-now standard form for quantization conditions in the RFT formalism, exemplified by \cref{eq:identQC}, aside from a sign difference arising from our boldface notation. All the complications introduced by spin are contained in the form of the building blocks $\textbf K_{\df,3}$, $\textbf G$, $\textbf F$, and $\textbf K_2$, which are discussed, respectively, in sections \ref{sec:Kdfsub}, \ref{sec:Gsub}, \ref{sec:Fsub}, and \ref{sec:K2sub}.

To use the quantization condition in practice, one must truncate the sum over angular momentum. This can be achieved, following ref.~\cite{\HSQCa}, by setting $\textbf K_2$ and $\textbf K_{\df,3}$ to zero for $\ell > \ell_{\sf max}$, since this has the effect of truncating also the contributions of $\textbf F$ and $\textbf G$. Such a truncation is reasonable near to threshold, where angular-momentum barriers suppress interactions for larger values of $\ell$. In principle, one can envisage using a different maximum angular momentum cutoff for $\textbf K_2$ and $\textbf K_{\df,3}$, although it is plausible that they should be related. In any case, since the sum over $\bm k$ is automatically truncated by the cutoff function, the truncation in $\ell$ leads to finite-dimensional matrices, and implementing the determinant condition becomes tractable.

A further practical step is to decompose the matrices into irreducible representations (irreps) of the appropriate finite-volume symmetry group, which allows the matrix in the quantization condition to be block diagonalized. The group depends on the total momentum $\boldsymbol P$. In the case of three neutrons, one must use fermionic irreps, and the construction of the projectors that block-diagonalize the matrices involves Wigner D matrices representing the effect of Wigner rotations. Details on irrep projection for the three-particle quantization condition for scalars have been discussed in refs.~\cite{\dwave,\HH}. Applying the same procedure for particles with spin should be straightforward.

\subsection{Relating \texorpdfstring{$\Kdf$}{the divergence-free K-matrix} to the scattering amplitude\label{sec:inteqs}}

The second step of the formalism provides a relation between the two- and three-particle K matrices, determined as outlined in the previous section, and the three-particle scattering amplitude, $\cM_3$. We obtain this relation using the correlator $\mathbf{M}_{3, L}^{(u, u)}$, whose decomposition is given in \Cref{eq:M3Ldef_3N} above. We first antisymmetrize the correlator to yield
\begin{equation}
\mathbf{M}_{3, L}(P) \equiv \mathcal{A}\left[\mathbf{M}_{3, L}^{(u, u)}(P)\right] \,.
\label{eq:M3L}
\end{equation}
The antisymmetrization step sums over all permutations of momenta on the external legs with appropriate signs and is described in detail in \Cref{app:symm}.

Second, we formally perform an infinitesimal shift in the total energy into the complex plane, denoted by $i \epsilon$, and then take an ordered double limit,
\begin{equation}
\mathbf{M}_{3}(E, \boldsymbol{P})=\lim _{\epsilon \rightarrow 0^{+}} \lim _{L \rightarrow \infty} \mathbf{M}_{3, L}(E+i \epsilon, \boldsymbol{P})\,.
\end{equation}
The quantity on the left-hand side is the full three-neutron scattering amplitude, defined as the sum of all diagrams in the skeleton expansion with a standard $i \epsilon$ pole prescription in propagators. The limit on the right-hand side yields a set of infinite-volume integral equations.

To spell this out in more detail, we begin with the two-to-two amplitude that appears as a subprocess of three-to-three scattering. We define this via the two-particle
K-matrix but introduce a version of the latter that is rescaled relative to that in \Cref{sec:K2sub}.
We denote this by
$\textbf K_2(\boldsymbol p)$, where $\boldsymbol p$ is the spectator momentum,
defining
\begin{equation}
    [\textbf K_2]_{ \ell'm' \bm m'^*_s ;  \ell m \bm m^*_s} (\boldsymbol p) \equiv i
    \mathcal K_2^{(\ell' m' \bm m'^*_{s}, \ell m \bm m^*_{s} )}(E_{2,p}^*) \,.
    \end{equation}
Comparing to eq.~\eqref{eq:boldK2_spin_def}, note that here we have dropped a factor of $2 \omega_p L^3$ and the Kronecker delta. We next define the corresponding phase-space factor
\begin{equation}
    [\boldsymbol \rho]_{ \ell'm' \bm m'^*_s ;  \ell m \bm m^*_s} (\boldsymbol p) = \delta_{\bm m'^*_s \bm m^*_s} \delta_{\ell' \ell} \delta_{m' m} \frac{ - i \vert q_{2,p}^* \vert \big (1 - H(\boldsymbol p) \big ) +  q_{2,p}^* }{16 \pi E_{2,p}^*} \,,
\end{equation}
such that the two-to-two scattering amplitude is given by
\begin{equation}
    \textbf M_{2}(\boldsymbol p) = \frac{1}{\textbf K_2(\boldsymbol p)^{-1} - \boldsymbol \rho(\boldsymbol p)} \,.
\end{equation}
Up to the rescaling by $2 \omega_p L^3$, this is the ordered double limit ($L \to \infty$ followed by $\epsilon \to 0^+$) of $\textbf M_{2,L}$, defined in eq.~\eqref{eq:M2Ldef}.

The next step is to evaluate the ordered limit for $\textbf D^{(u,u)}$; we denote the result of this by $\textbf D_\infty^{(u,u)}(\boldsymbol p, \boldsymbol k)$, where the spectator-momentum indices have been promoted to arguments of the function. As described, for example,
in eq.~(85) of ref.~\cite{Hansen:2015zga}, the result of the limit can be expressed via an integral equation
\begin{equation}
    \textbf D_\infty^{(u,u)}(\boldsymbol p, \boldsymbol k)  \equiv   \textbf M_{2}(\boldsymbol p) \textbf G_\infty(\boldsymbol p, \boldsymbol k) \textbf M_{2}(\boldsymbol k) + \textbf M_{2}(\boldsymbol p)   \int_{\boldsymbol k'} \textbf G_\infty(\boldsymbol p, \boldsymbol k') \textbf D_\infty^{(u,u)}(\boldsymbol k', \boldsymbol k) \,,
\end{equation}
where $\int_{\boldsymbol k} = \int d^3 \boldsymbol k[(2 \pi)^3(2 \omega_k)]^{-1}$,
 and $\textbf G_\infty(\boldsymbol p, \boldsymbol k)$,
 the exchange factor appropriate for the integral equation, is given by
\begin{equation}
    [\textbf G_{\infty}]_{\ell' m' \bm m'^*_s;  \ell m \bm m^*_s}(\boldsymbol p, \boldsymbol k) = 4 \omega_p \omega_k L^6 \, \textbf G_{p \ell' m' \bm m'^*_s; k \ell m \bm m^*_s} \,,
\end{equation}
which is just a rescaled version of the finite-volume analog with the spectator-momentum indices promoted to arguments of the function. Note that, although the convention for $\int_{\boldsymbol k}$ differs as compared to ref.~\cite{Hansen:2015zga}, this is compensated by differences elsewhere such that the convention for $\textbf D_\infty^{(u,u)}(\boldsymbol p, \boldsymbol k)$ is the same.

Once $\textbf D_\infty^{(u,u)}(\boldsymbol p, \boldsymbol k)$ is determined via the integral equation, one can readily evaluate the left and right dressing functions via
\begin{align}
    \textbf L^{(u)}_\infty(\boldsymbol p, \boldsymbol k) & = \left [ \frac{1}{3} + \textbf M_{2}(\boldsymbol k) \boldsymbol \rho(\boldsymbol k) \right   ]  \tilde \delta(\boldsymbol p - \boldsymbol k) +   \textbf D_\infty^{(u,u)}(\boldsymbol p, \boldsymbol k) \boldsymbol \rho(\boldsymbol k)
    \\
    \textbf R^{(u)}_\infty(\boldsymbol p, \boldsymbol k) & =  \left [ \frac{1}{3} + \boldsymbol \rho(\boldsymbol k) \textbf M_{2}(\boldsymbol k)  \right   ]   \tilde \delta(\boldsymbol p - \boldsymbol k)  + \boldsymbol \rho(\boldsymbol p) \textbf D_\infty^{(u,u)}(\boldsymbol p, \boldsymbol k) \,,
\end{align}
where $\tilde \delta(\boldsymbol p - \boldsymbol k) = 2 \omega_p (2 \pi)^3 \delta^3(\boldsymbol p - \boldsymbol k)$. [See eqs.~(92) and (94) of ref.~\cite{Hansen:2015zga}. However, for these quantities our convention differs by factors of $2 \omega$ from that work.]
Either of these can then be used to construct a final integral equation for a quantity denoted by $ \textbf T(\boldsymbol p, \boldsymbol k)$ that sums an infinite series of $\textbf K_{\text{df},3}$ insertions:
\begin{equation}
    \textbf T(\boldsymbol p, \boldsymbol k) = \textbf K_{\text{df},3}(\boldsymbol p, \boldsymbol k) + \int_{\boldsymbol p'} \int_{\boldsymbol k'} \textbf K_{\text{df},3}(\boldsymbol p, \boldsymbol p') \boldsymbol \rho(\boldsymbol p') \textbf L^{(u)}_\infty(\boldsymbol p', \boldsymbol k')   \textbf T(\boldsymbol k', \boldsymbol k) \,.
\end{equation}
[See also eq.~(91) of ref.~\cite{Hansen:2015zga}, where the convention here matches without rescaling.]
This differs from the final amplitude because it is missing contributions from pairwise scattering outside the $\textbf K_{\text{df},3}$ factors and in the $\textbf K_{\text{df},3}$-independent term. Restoring these contributions and anti-symmetrizing gives
\begin{equation}
\textbf M_3(E, \boldsymbol P) = \mathcal A \left [  \textbf D_\infty^{(u,u)}(\boldsymbol p, \boldsymbol k) +  \int_{\boldsymbol p'} \int_{\boldsymbol k'}  \textbf L^{(u)}_\infty(\boldsymbol p, \boldsymbol p')   \textbf T(\boldsymbol p', \boldsymbol k')  \textbf R^{(u)}_\infty(\boldsymbol k', \boldsymbol k)\right ] \,.
\end{equation}
This is the final result for the three-neutron scattering amplitude.

%
Dedicated studies concerning the solution to the corresponding integral equations
for scalar particles can be found in refs.~\cite{Jackura:2020bsk,Dawid:2023jrj}, and we expect many of the practical details to carry over to the three-neutron case.

\subsection{Change of basis and specific truncations\label{sec:changeofbasis}}

In this subsection we consider additional practical aspects of implementing the quantization condition, focusing on those that are particular to the presence of spin degrees of freedom.

We first note that it will be useful, in many cases, to rotate all matrices to the basis in which the dimer has definite total spin (as defined in the dimer-axis frame). This total-spin basis is already discussed in the context of the two-particle K-matrix, $\textbf K_2$, in \Cref{sec:K2sub}, in particular in \Cref{eq:K2totals}. Extending the notation introduced there, we write
\begin{multline}
\Big \vert \boldsymbol k, \ell, m, s^*, \mu^*_s, m_s(\klittle) \Big \rangle \\
= \sum_{ m'_{s}(\boldsymbol a^*) , m'_{s}(\boldsymbol b^*) } \Big \vert \boldsymbol k, \ell, m, m'_{s}(\astarlittle) , m'_{s}(\bstarlittle) , m_s(\klittle) \Big \rangle
\Big \langle \nicefrac12 \ m'_{s}(\astarlittle) , \ \nicefrac12 \ m'_{s}(\bstarlittle)
\Big \vert s^*, \mu^*_s
\Big \rangle
\,,
\label{eq:spinbasis}
\end{multline}
where $s^* \in \{ 0, 1\}$ is the total dimer spin and $\mu^*_s$ is the corresponding azimuthal component. Note that we do not assign any momentum label to these spin indices as the $*$ is sufficient to denote the dimer-axis frame. The last factor in \Cref{eq:spinbasis} is a standard $SU(2)$ Clebsch-Gordon coefficient for the reduction $ \nicefrac12 \otimes \nicefrac12 = 0 \oplus 1$.

This change of basis block diagonalizes $\textbf K_2$ to a spin-zero and spin-one block and leaves $\textbf F$ proportional to the identity matrix in spin-space, with index structure $\delta_{ s^{\prime*}, s^*} \ \delta_{\mu_s^{\prime*} , \mu^*_s }$. By contrast, $\textbf G$ and $\textbf K_{\text{df}, 3}$ are not expected to have any simplifications from the rotation to this basis.

Similarly, it is possible to perform a change of basis to combine $s^*$ and $\ell$ into a total two-particle angular momentum $j$. To do this in practice one needs to fix a particular value of $\ell_{\sf{max}}$, leading to a finite set of $\ell$ values that can be combined with each $s^*$ to reach a finite set of states labelled by $\boldsymbol k, j, \mu_j, \ell, s^*, m_s(\klittle)$. For the case of $\ell_{\sf{max}} = 0$, for example, one has only the channel $\{ j, \ell, s^* \} = \{0, 0, 0 \} $, since $s^*=1$ requires odd values of $\ell$. For $\ell_{\sf{max}} = 1$, we add an $s^*=1$ component, so that there are now several channels
\begin{equation}
\{ j, \ell, s^* \} \in \big \{ \{0, 0, 0 \}, \{0, 1, 1 \}, \{ 1, 1, 1 \} , \{ 2, 1, 1 \} \big \} \,.
\end{equation}
In this case no mixing occurs in $\textbf K_2$ (since the two $j=0$ cases have opposite parities), but $\textbf K_{\text{df},3}$ and $\textbf G$ will mix the channels, since $j$ is not a good quantum number of the three-particle state. Mixing in the two-particle K-matrix first arises for $\ell_{\sf{max}} = 3$, as discussed in \Cref{sec:K2sub}.

\section{Parametrization of \texorpdfstring{$\Kdf$}{the divergence-free K-matrix}\label{sec:Kdf}}

Implementing the quantization condition for three identical fermions requires a para\-metrization of $\kdf$. Since $\kdf$ is, by construction, smooth aside from three-particle resonances or bound states, we present its parametrization in an expansion about threshold. For this expansion, we work with the quantity $\mathcal K_\text{df,3}^{\sf{lab}}$ of \Cref{eq:all_arguments}, in which spin components are defined with respect to the lab-frame axis. When using $\Kdf$ in, e.g., the evaluation of the quantization condition, it will be necessary to rotate the spin indices to the dimer-frame axis using \Cref{eq:basisspinrotation}. The key constraint is that $\mathcal K_\text{df,3}^{\sf{lab}}$ must have the same symmetry properties as the scattering amplitude, $\cM_3$. It must transform covariantly under Lorentz transformations and parity, and be fully antisymmetric with respect to the simultaneous exchange of the spin and momentum labels for any pair of incoming or outgoing particles.

To construct the threshold expansion, we proceed by writing down Lorentz- and parity-invariant local operators composed of three-neutron fields and three conjugate fields and their derivatives. These have the schematic form $ \overline {\mathcal N}^3 {\mathcal N}^3$ (with derivatives and Dirac indices implicit). By using quantum fields ${\mathcal N}$ and $\overline {\mathcal N}$, we automatically enforce the required antisymmetry property, and by using local operators we ensure that the momentum dependence is smooth. If we enumerate all possible independent operators with up to a certain number of derivatives, each multiplied by an independent coefficient, then, by the standard assumption of effective field theories (EFTs), the corresponding matrix elements will yield the most general amplitudes consistent with the symmetries.

To convert the local operators into explicit forms for $\mathcal K_\text{df,3}^{\sf{lab}}$, we take matrix elements of the operators between external states in the lab-frame basis. This leads to expressions involving the Dirac spinors associated with the initial and final particles, because of the relations
\begin{equation}
\langle \bm k', m_s(\kprimelittle) | \overline {\mathcal N}(x) | 0 \rangle = \bar u_{m_s(\bm k')}(\bm k') e^{ i k' \cdot x}\,,
\qquad
\langle 0 | {\mathcal N}(x) | \bm k, m_s(\bm k) \rangle = u_{m_s(\bm k)}(\bm k) e^{-i k\cdot x}\,.
\end{equation}
In this way an operator such as
\begin{equation}
\cO_{\rm SSS}(x) = [\overline {\mathcal N}(x) {\mathcal N}(x)]^3 \,,
\end{equation}
leads to an amplitude
\begin{multline}
\langle \bm k', m_s(\kprimelittle); \bm a',m_s(\aprimelittle); \bm b',m_s(\bprimelittle) | \cO_{\rm SSS}(x) |
\bm k, m_s(\bm k); \bm a, m_s(\bm a); \bm b,m_s(\bm b) \rangle = \\
6 (\bar u_{k'} u_k) (\bar u_{a'} u_a) (\bar u_{b'} u_b)
- 6 (\bar u_{k'} u_a) (\bar u_{a'} u_k) (\bar u_{b'} u_b) + \dots\,.
\label{eq:SSSdecomp}
\end{multline}
Here we are using the shorthand $u_k = u_{m_s(\bm k)} (\bm k)$, etc., and the ellipsis represents the four other possible permutations arising from Wick contractions. We are also assuming that the external states are chosen to satisfy conservation of four-momentum. A key point here is that Dirac spinors describe spin components in the lab-frame basis. The right-hand side of \Cref{eq:SSSdecomp} has the correct symmetry properties to be a contribution to $\mathcal K_\text{df,3}^{\sf{lab}}$: it is manifestly antisymmetric under initial or final particle interchanges, and it inherits the correct Lorentz and parity transformation properties from those of the free-particle states.

We now consider the complete set of operators without derivatives, that is, with dimension of $[E]^9$. The available Lorentz- and parity-invariant choices are $\cO_{\rm SSS}$ and
\begin{align}
\begin{split}
\cO_{\rm SPP} &= [\overline {\mathcal N} {\mathcal N}] [\overline {\mathcal N} \gamma_5 {\mathcal N}] [\overline {\mathcal N} \gamma_5 {\mathcal N}]\,,
\\
\cO_{\rm SVV} &= [\overline {\mathcal N} {\mathcal N}] [\overline {\mathcal N} \gamma_\mu {\mathcal N}] [\overline {\mathcal N} \gamma^\mu {\mathcal N}]\,,
\\
\cO_{\rm SAA} &= [\overline {\mathcal N} {\mathcal N}] [\overline {\mathcal N} \gamma_\mu\gamma_5 {\mathcal N}] [\overline {\mathcal N} \gamma^\mu\gamma_5 {\mathcal N}]\,,
\\
\cO_{\rm STT} &= [\overline {\mathcal N} {\mathcal N}] [\overline {\mathcal N} \sigma_{\mu\nu} {\mathcal N}] [\overline {\mathcal N} \sigma^{\mu\nu} {\mathcal N}]\,,
\\
\cO_{\rm PVA} &= [\overline {\mathcal N} \gamma_5 {\mathcal N}] [\overline {\mathcal N} \gamma_\mu {\mathcal N}] [\overline {\mathcal N} \gamma^\mu\gamma_5 {\mathcal N}]\,,
\\
\cO_{\rm PTT'} &= [\overline {\mathcal N} \gamma_5 {\mathcal N}] [\overline {\mathcal N} \sigma_{\mu\nu} {\mathcal N}]
[\overline {\mathcal N} \sigma^{\mu\nu}\gamma_5 {\mathcal N}]\,,
\\
\cO_{\rm TVV} &= [\overline {\mathcal N} \sigma_{\mu\nu} {\mathcal N}] [\overline {\mathcal N} \gamma^\mu {\mathcal N}] [\overline {\mathcal N} \gamma^\nu {\mathcal N}]\,,
\\
\cO_{\rm TAA} &= [\overline {\mathcal N} \sigma_{\mu\nu} {\mathcal N}] [\overline {\mathcal N} \gamma^\mu\gamma_5 {\mathcal N}]
[\overline {\mathcal N} \gamma^\nu\gamma_5 {\mathcal N}]\,,
\\
\cO_{\rm T'VA} &= [\overline {\mathcal N} \sigma_{\mu\nu}\gamma_5 {\mathcal N}] [\overline {\mathcal N} \gamma^\mu {\mathcal N}]
[\overline {\mathcal N} \gamma^\nu\gamma_5 {\mathcal N}]\,,
\\
\cO_{\rm TTT} &= [\overline {\mathcal N}\sigma_{\mu\nu} {\mathcal N}] [\overline {\mathcal N} \sigma^{\nu\rho} {\mathcal N}]
[\overline {\mathcal N} {\sigma_{\rho}}^\mu {\mathcal N}]\,,
\\
\cO_{\rm TT'T'} &= [\overline {\mathcal N}\sigma_{\mu\nu} {\mathcal N}] [\overline {\mathcal N} \sigma^{\nu\rho}\gamma_5 {\mathcal N}]
[\overline {\mathcal N} {\sigma_{\rho}}^\mu\gamma_5 {\mathcal N}]\,,
\label{eq:ops1}
\end{split}
\end{align}
where we have left the $x$ arguments implicit, and used
\begin{equation}
\sigma_{\mu\nu} = \tfrac{i}{2} \left[\gamma_\mu,\gamma_\nu\right]\,.
\end{equation}
This list can be shortened using Fierz identities, but we give a complete enumeration as this would be relevant if we were to consider nonidentical fermions. We convert these operators into forms for $\mathcal K_\text{df,3}^{\sf{lab}}$ by taking matrix elements as in \Cref{eq:SSSdecomp}. We then explicitly evaluate the momentum dependence using {\tt Mathematica} for arbitrary choices of spinor components (not yet constrained to satisfy the Dirac equation) and find that all 12 operators lead to forms that are proportional.%
\footnote{%
By contrast, there are three ``four-fermion'' Lorentz- and parity-operators of the form $\overline {\mathcal N}^2 {\mathcal N}^2$. The presence of only one ``six-fermion'' operator is expected at quadratic order in a nonrelativistic expansion, where it is given by \Cref{eq:O1}. What is surprising is that this holds to all orders in this expansion.
}
Thus, with zero derivatives, there is a single contribution to $\mathcal K_\text{df,3}^{\sf{lab}}$, given by the right-hand side of \Cref{eq:SSSdecomp}.

We now enforce that the spinors satisfy the Dirac equation by writing
\begin{equation}
u_k = \sqrt{2 \omega_{ k}}
\begin{pmatrix}
\chi_k \\
\frac{\boldsymbol{\sigma} \cdot \bm k}{\omega_{ k} + m} \chi_k
\end{pmatrix}\,,
\label{eq:spinor}
\end{equation}
where $\chi_k$ is the non-relativistic two-spinor corresponding to the component $m_s(\bm k)$, and $\omega_{ k}=\sqrt{\bm k^2+m^2}$. We insert this into \cref{eq:SSSdecomp}, and perform a nonrelativistic expansion, i.e.~an expansion in powers of $\bm k/m$. The leading-order term (with no factors of $\bm k$) vanishes, as is expected because one cannot antisymmetrize the spin wavefunction of three identical spin-$1/2$ particles. The first nonvanishing term is quadratic in momenta and proportional to
\begin{align}
\cK_{A} & = \, \overline\cA \left[ (\chi_{k'}^\dagger \: \boldsymbol{\sigma} \cdot \boldsymbol{k'} \:
\boldsymbol{\sigma} \cdot \boldsymbol{k} \: \chi_k)(\chi_{a'}^\dagger \chi_a)(\chi_{b'}^\dagger \chi_b) \right] \,.
\label{eq:O1}
\end{align}
Here $\overline \cA$ indicates antisymmetrization over initial and final particle labels; it differs from the operation $\cA$ defined in \Cref{eq:antisymmetrizedef}, and used in \Cref{eq:M3L}, by not requiring an initial step of combining with spherical harmonics. Therefore, the contribution to $\kdf$ is:
\begin{equation}
m^2 \kdf^{\sf lab} \supset \frac{ c_0 }{m^2} \cK_{A} + O\left( \frac{\bm k^4}{m^4} \right),
\label{eq:kdfE9}
\end{equation}
where $c_0$ is a dimensionless coefficient whose value is not fixed. Here we assume that the contribution of the operators in \Cref{eq:ops1} appears in the Lagrangian as ${\mathcal L \supset (g_0/m^5) \cO}$, where $g_0$ is dimensionless and proportional to $c_0$ in \Cref{eq:kdfE9}. The expansion can be continued to higher orders in $\bm k/m$, a point that we return to below.

At quadratic order in the nonrelativistic expansion, we must also consider operators containing two derivatives,%
\footnote{%
Operators with one derivative can be related to operators without derivatives using the equations of motion, and thus are not independent.
}
i.e. of energy dimension $[E]^{11}$. In \Cref{app:twoderivs} we enumerate all such operators that are Lorentz- and parity-invariant, that are not related by Fierz identities, and that cannot be written in terms of operators without derivatives using the equations of motion. We find 22 such operators. However, if we insert the nonrelativistic expansion of the spinors into the corresponding contributions to $\mathcal K_\text{df,3}^{\sf{lab}}$, we find only two independent terms at quadratic order,%
\footnote{%
Because the zeroth component of the derivatives yield energies, which do not vanish at threshold, there could in principle be a zeroth order term, but this vanishes due to the antisymmetry, as discussed above.
}
which can be chosen to be $\cK_{A}$, given above, and
\begin{align}
\cK_{B} & = \overline\cA \left[ \boldsymbol{k'} \cdot \boldsymbol{k}(\chi_{k'}^\dagger \chi_k)(\chi_{a'}^\dagger \chi_a)(\chi_{b'}^\dagger \chi_b) \right] \,.
\end{align}
Since these operators have a higher energy dimension, the couplings in the Lagrangian of the EFT will contain two inverse powers of the typical energy scale of the EFT, $\Lambda^2_\text{EFT}$. This way, the contribution of dimension-11 operators to $\kdf$ at quadratic order in momenta is:
\begin{equation}
m^2 \kdf^{\sf lab} \supset \frac{c_1}{\Lambda^2_\text{EFT}} \cK_{A} +\frac{c_2}{\Lambda^2_\text{EFT}} \cK_{B} \, + O\left( \frac{\bm k^4}{m^2 \Lambda^2_\text{EFT}} \right).
\label{eq:kdfE11}
\end{equation}
If we consider systems of nucleons at low momentum described by pionless EFT, the energy scale is the pion mass, $\Lambda_\text{EFT} \simeq m_\pi$. This choice implies that (i) the contribution from \Cref{eq:kdfE9} is subdominant with respect to that of \Cref{eq:kdfE11}, and (ii) \Cref{eq:kdfE11} is the most general form of $\kdf$ through $\cO(\bm k^2)$. We note, however, that the range of convergence of pionless EFT is limited by the left-hand cut ($\bm k^2<m_\pi^2/4$), while the relativistic finite-volume three-neutron formalism is applicable beyond that, i.e.~up to the $NNN\pi$ threshold ($\bm k^2 \sim m_\pi m_N$). We also stress that, at this order in the nonrelativistic expansion, we would have obtained the same result simply by enforcing rotation and parity invariance and antisymmetry.

Proceeding to higher order in the nonrelativistic expansion is straightforward in principle. There are no terms of cubic order, due to the requirement of parity invariance of $\kdf$. This is because momenta flip sign, while spins remain unchanged. To determine the independent terms of quartic order would require the enumeration of allowed operators with up to four derivatives, a straightforward but tedious task that we have not undertaken. We stress that an approach in which one simply writes down operators consistent with rotation and parity invariance and antisymmetry, while working at quadratic order, would not, in general, work at higher orders, because constraints due to the Lorentz covariance would be lost (just as the relative size of the $\bm k^2/m^2$ and $\bm k^4/m^4$ terms in the relativistic dispersion relation cannot be determined without enforcing Lorentz symmetry).

As a final comment, we note that there is a superficial difference between the nonrelativistic expansion that we have used here, and the form of the threshold expansions for mesonic systems (e.g.~three pions in refs.~\cite{\dwave,\isospin}). The latter are written in terms of Lorentz invariant Mandelstam variables, whereas here we explicitly expand in powers of the nucleon's three-momentum, $\bm k$. There is, however, no fundamental difference between the approaches, because the quantities in the mesonic case can be rewritten in terms of a nonrelativistic expansion. For example, the quantity $\Delta=(s-9m^2)/(9m^2)$ used in ref.~\cite{\dwave} as an expansion parameter, is proportional to the sum of $\bm k^2$ for the initial and final particles, plus higher order terms. It just turns out that, in the absence of the spin degrees of freedom, one can write the expansion in a manifestly Lorentz-invariant way more easily than in the present case.

\section{Conclusions}
\label{sec:conc}

This work presents the finite-volume formalism in the RFT approach for three identical spin-1/2 particles. The most relevant physical system to which it is applicable is three neutrons (or protons), but also, e.g., three spin-1/2 hyperons. The new features are the presence of the spin degrees of freedom and the overall antisymmetry of the states. Since the formalism is relativistic, spin components can mix via boosts, which is indeed the main technical complication of the derivation. The solution is to include the effect of Wigner rotations when relating spin degrees of freedom in the lab (finite-volume) frame and the CMF of pairs---see \Cref{sec:relspin}.

Sections \ref{sec:ModOfBB} and \ref{sec:derivation} represent the core of this work. First, the modifications of the building blocks of the three-particle formalism are discussed in \Cref{sec:ModOfBB}. Defining the spin degrees of freedom of the interacting pair in the dimer frame leads to very simple generalizations of the blocks $\textbf F$ and $\textbf K_2$, which are related to sum-minus integral difference of a two-neutron loop and two-neutron interactions, respectively. By contrast, in the dimer frame, the fermionic nature of the particles is explicit in $\textbf G$---the builing block related to one-neutron exchange processes. As such, $\textbf G$ contains the aforementioned Wigner rotations and minus sign from antisymmetry (see \Cref{eq:Gspin}). In \Cref{sec:derivation}, the derivation of the finite-volume formalism using a diagrammatic expansion of the finite-volume correlator that utilizes those building blocks is presented. Simple examples of contributions to this expansion are depicted in \Cref{fig:deri}.

The outcomes of \Cref{sec:QC} are the finite-volume correlator in \Cref{eq:CLresult_3N} and the finite-volume analogous of the (unsymmetrized) scattering amplitude in \Cref{eq:M3Ldef_3N}. The quantization condition and the integral equations relating the three-particle K matrix to the scattering amplitude are easily obtained from these equations. They can be found in \Cref{sec:summary}, where we summarize the results of the derivation. As noted, we have focused on the derivation and form of the resulting quantization in this work, providing only a sketch of many of the details of the implementation of the formalism. We will discuss these in a follow-up work.

An important ingredient in the finite-volume formalism is the parametrization of the three-neutron K matrix, which describes short-range three-neutron interactions. In \Cref{sec:Kdf} we show, working to lowest nontrivial order in a nonrelativistic momentum expansion, that only two different types of operators contribute---see \Cref{eq:kdfE11}. At higher orders, more independent terms appear, as discussed in \Cref{app:twoderivs}.

Now that we have determined how to include spin into the RFT approach in the simplest setting of identical fermions, there are several generalizations that can be considered. These include three nucleons of arbitrary isospin (in isosymmetric QCD), which should involve a straightforward combination of the present work with the formalism developed for the generic three-pion system~\cite{\isospin}. Other systems of interest involve two spin-1/2 particles and a (pseudo-)scalar, or one spin-1/2 particle and two (pseudo-)scalars, for which a combination of the methods presented here with those for nondegenerate systems~\cite{\BSnondegen,\BStwoplusone} will be needed. A particularly interesting example in QCD is the $N\pi\pi$ system at non-maximal isospin, which has the additional difficulty of mixing with the $N\pi$ state. In the context of the RFT approach, this will require as well the use of a generalization of the $2\leftrightarrow 3$ formalism of ref.~\cite{\BHSQC}. Such a generalization is of great interest, as it will allow access to the Roper resonance, a puzzling baryonic excitation with the same quantum numbers of the nucleon.

\acknowledgments

We thank Ra{\'u}l Brice{\~n}o, Vincenzo Cirigliano, Will Detmold, Dan Hackett, Andrew Jackura, and Felix Ziegler for useful discussions.

MTH is supported by UKRI Future Leader Fellowship MR/T019956/1 and in part by UK STFC grant ST/P000630/1. The work of FRL has been supported in part by the U.S.~Department of Energy, Office of Science, Office of Nuclear Physics, under grant Contract Numbers DE-SC0011090 and DE-SC0021006. FRL acknowledges financial support by the Mauricio and Carlota Botton Fellowship. The work of SRS and ZTD is supported in part by the USDOE grant No.~DE-SC0011637.

FRL would like to thank the Physics Department at the University of Washington for its hospitality during a visit in which this work was initiated.

\appendix

\section{Further details of the derivation}
\label{app:deriv}

In this appendix we provide additional details of the derivation sketched in the main text in \Cref{sec:derivation}. In particular, we discuss the changes that must be made compared to the derivation for identical scalars, given in ref.~\cite{\HSQCa} (referred to as HS throughout this appendix) due to the antisymmetric nature of the three-neutron state. In this regard, it is important to keep in mind that the antisymmetry of intermediate states is not fully manifested in the individual kinematic matrices $F$ and $G$ that appear in the quantization condition. Instead, these quantities can implement the antisymmetry of the nonspectator pair%
\footnote{%
Strictly speaking it is a choice whether to implement the antisymmetry in $F$ and $G$, as these are kinematic quantities that could be defined independently of the symmetry of the particles involved. The antisymmetry then arises because $F$ and $G$ appear next to amplitudes, such as $\cK_2$ and $\Kdf$, that do have the requisite (anti)symmetry, and thus project out the parts of $F$ and $G$ with the corresponding symmetry. We choose to build this projection into $F$ and $G$ from the start.
}
by restrictions on allowed values of $s$ and $\ell$. Full antisymmetry is only recovered when combining $F$ and $G$, and for this the overall minus sign in $G$ is essential.

A closely related observation is that the antisymmetry is explicitly broken by the variables we choose, since we single out one of the particles to be the spectator. This leads to the presence, at intermediate stages of the derivation, of on-shell three-particle amplitudes that are not fully antisymmetric under particle interchange. The archetypical such quantity in HS is $\mathcal K_{\text{df},3}^{(u,u)}$, the unsymmetrized form of $\Kdf$. Its lack of symmetry is analogous to that for the quantity $\textbf M_{3,L}^{(u,u)}$ [see \Cref{eq:M3Ldef_3N}]. In HS, these quantities are denoted by superscripts $(u)$, $(s)$ and $(\tilde s)$, and we follow the same notation here. These labels describe the attachment of the spectator momentum to the diagrams under consideration. We denote the spectator momentum by $k$, and the pair momenta by $p$ and $b$. If the final interaction before the external momenta involves two particles, then the $(u)$, $(s)$ and $(\tilde s)$ quantities, respectively, are those in which the spectator particle has momentum $k$, $p$ and $b$. (See figure~13(b) of HS---``$u$" stands for unscattered and ``$s$" for scattered.) If the final interaction involves all three particles, then the contribution is divided equally into the three types of quantities.

The previous description works for identical bosons, but does not provide a complete prescription for fermions, due to the antisymmetry of the nonspectator pair. In particular, one must choose one of the pair to be ``privileged'', in the sense that, in some convention, it is the particle with respect to which the antisymmetry is defined. Our choice for the privileged particles are that they have momenta $p$, $b$ and $k$, respectively, for the $(u)$, $(s)$, and $(\tilde s)$ amplitudes. Another way of describing this is to list the momenta such that the first is that of the spectator, the second that of the privileged member of the pair, and third that of the other member of the pair. Then the ordering is $\{k,p,b\}$ for $(u)$, $\{p,b,k\}$ for $(s)$, and $\{b,k,p\}$ for $(\tilde s)$, (i.e. the cyclic permutations of the $(u)$ ordering). With this choice, a fully {\em antisymmetric} amplitude is given by {\em adding} the three terms
\begin{equation}
A = A^{(u)} + A^{(s)} + A^{(\tilde s)}\,,
\label{eq:Asym}
\end{equation}
a result that, perhaps counterintuitively, takes the same form as that for identical bosons. This is shown explicitly in the example described in \Cref{app:symm}. With this choice, we can carry over most of the equations from the derivation in HS with minimal changes.

In the following, we assume that the reader has a copy of HS in front of them, and refer to equations in that work as (HS1), (HS2), etc. We do not repeat these equations here.

Among the most challenging parts of the derivation in HS is showing how asymmetric quantities such as $A^{(u)}$ end being symmetrized in the final result. Here the same issue arises, but involves antisymmetrization. (Note that for amplitudes with initial and final momenta, like $\textbf K_{\df,3}$, one must antisymmetrize separately the initial and final momenta.) Various identities are needed for this to work, and these must be generalized here from the versions applicable to identical bosons given in HS. The first example of such generalizations is in the definitions of $(u)$, $(s)$ and $(\tilde s)$ already described above, i.e. with the choice of a privileged momentum. The next example concerns the discussion between (HS140) and (HS146), which notes that the combination $A^{(u)}+2 A^{( s)}$ appears frequently in the derivation, rather than the desired combination of \Cref{eq:Asym}. It is then explained that these two quantities are, in fact, equal, given the projection onto even angular momenta that occurs for identical bosons. The same conclusion holds here, although the argumentation is slightly changed, as we now describe.

In particular, a key result is that $(s)$ amplitudes always arise in the derivation connected to $(u)$ amplitudes by an $\mathbf F$ matrix, an example being $\mathbf A_3'^{(s)} \mathbf F \mathbf A_3^{(u)}$. Now, $(u)$ amplitudes are antisymmetric under the interchange of the particles in the pair, leading to the constraints on $\ell$ and $s$ described in the main text. This antisymmetry is preserved by $\mathbf F$, because it commutes with $(-1)^\ell$ (up to exponentially-suppressed corrections, a result derived in HS Appendix A), and also commutes with projectors onto definite pair spin (given the trivial nature of the spin index structure of $\mathbf F$). Thus, the only parts of the $(s)$ amplitudes that contribute are those that are antisymmetric under pair exchange. These parts are, however, identical to the corresponding parts of the $(\tilde s)$ amplitudes. This is because to go from the $\{pkb\}$ ordering for $(s)$ to the $\{bpk\}$ ordering for $(\tilde s)$, one must both interchange $b$ and $p$, which leads to a minus sign due to the above-described antisymmetric projection, {\rm and} interchange $p$ and $k$, which leads to a second minus sign as these are now the momenta associated with the interacting pair, and the amplitude is antisymmetric under their exchange. Because of this identity, we can freely exchange all appearances of $2 \mathbf A_3^{(s)}$ (for example) with $\mathbf A_3^{(s)}+\mathbf A_3^{(\tilde s)}$.

A second place where the argumentation of HS needs to be changed concerns the results in (HS196)-(HS198), which show that combinations like $\mathbf A_3' \mathbf F (\mathbf A_3^{(u)} - \mathbf A_3^{(s)})$, which appears to be a finite-volume term, is, in fact, an infinite-volume quantity. This result is used repeatedly in HS in the steps that lead to the recasting of the quantization condition in terms of quantities symmetric under particle exchange. Here it turns out that conclusion of (HS196)-(HS198) remains valid, but this now follows because $\mathbf A_3'$ is {\em antisymmetric} under the exchange of the two on-shell momentum labels (which would be $k$ and $p$ in the notation above), while $(\mathbf A_3^{(u)}-\mathbf A_3^{(s)})$ is {\em symmetric} [because the $k\leftrightarrow p$ interchange sends the $\{k,p,b\}$ ordering for $(u)$ into $\{p,k,b\}$, which leads to $ \mathbf A_3^{(u)} \to -\mathbf A_3^{(s)}$, since the latter is defined with the ordering $\{p,b,k\}$]. These symmetry properties are opposite to those in HS, but all that matters for the argument in HS to go through is that the symmetry properties of the quantities on the left and right of $\mathbf F$ are opposite, and this still holds.

With these observations, all the steps in the derivation of HS go through unchanged, except for the inclusion of spin-related factors as described in the main text. Thus, the form of the resulting quantization condition is unchanged (modulo the absorption of various factors into boldfaced quantities).

\section{Antisymmetrizing the scattering amplitude\label{app:symm}}

In this appendix we explain in detail the symmetrization and antisymmetrization procedures referred to in the main text---see, in particular, \Cref{sec:recap,sec:inteqs} and \Cref{app:deriv}.

We first recall how {\em symmetrization} works in the case of identical scalars~\cite{\HSQCa,\HSQCb}. This operation has been used in \Cref{eq:symmetrizeM3}. It acts on an unsymmetrized finite-volume quantity defined using our standard matrix indices, generically called $X_{k^{\prime} \ell^{\prime} m^{\prime}, k \ell m}^{(u, u)}$ (with the subscript $L$ dropped for brevity), and is defined by
\begin{align}
X\left(\boldsymbol{k}^{\prime}, \boldsymbol{a}^{\prime}, \boldsymbol{b}^{\prime} |
\boldsymbol{k}, \boldsymbol{a}, \boldsymbol{b}\right)
&\equiv \mathcal{S}\left[X_{k^{\prime} \ell^{\prime} m^{\prime}, k \ell m}^{(u, u)}\right]
\\
&\equiv
\sum_{
\{\boldsymbol{p}_{3}^{\prime}, \boldsymbol{p}_{1}^{\prime}\} \in \mathcal{P}_{3}^{\prime} }
\sum_{
\{\boldsymbol{p}_{3}, \boldsymbol{p}_{1}\} \in \mathcal{P}_{3}
}
X^{(u, u)}\left(\boldsymbol{p}_{3}^{\prime}, \boldsymbol{p}_{1}^{\prime} | \boldsymbol{p}_{3}, \boldsymbol{p}_{1}\right)\,,
\end{align}
where we combine the matrix amplitude with spherical harmonics to obtain the full momentum dependence,
\begin{equation}
X^{(u, u)}\left(\boldsymbol{k}^{\prime}, \boldsymbol{a}^{\prime} | \boldsymbol{k}, \boldsymbol{a}\right) \equiv 4 \pi
Y_{\ell^{\prime} m^{\prime}}^{*}(\hat{\boldsymbol{a}}_{2, k^{\prime}}^{\prime *})
X_{k^{\prime} \ell^{\prime} m^{\prime}, k \ell m}^{(u, u)} Y_{\ell m}(\hat{\boldsymbol{a}}_{2, k}^{*})\,,
\end{equation}
and the sets of permuted momenta are
\begin{equation}
\mathcal{P}_{3}=\{\{\boldsymbol{k}, \boldsymbol{a}\},\{\boldsymbol{a}, \boldsymbol{b}\},\{\boldsymbol{b}, \boldsymbol{k}\}\} \text { and } \mathcal{P}_{3}^{\prime}=\left\{\left\{\boldsymbol{k}^{\prime}, \boldsymbol{a}^{\prime}\right\}, \left\{\boldsymbol{a}^{\prime}, \boldsymbol{b}^{\prime}\right\}, \left\{\boldsymbol{b}^{\prime}, \boldsymbol{k}^{\prime}\right\}\right\} \,.
\end{equation}
Although we can take $\bm k'$ and $\bm k$ to lie in the finite-volume set, the constraint of having three on-shell particles will not, in general, allow the other external momenta ($\bm a'$, \dots) to lie in this set. Thus, to apply this definition, we must use the freedom to extend the definitions of finite-volume quantities to momenta lying outside the finite-volume set, which has been discussed in ref.~\cite{\HSQCb}. This issue becomes irrelevant in the $L\to\infty$ limit, which is what we ultimately take in most applications of (anti-)symmetrization.

Turning now to identical fermions, a generic asymmetric finite-volume quantity has a similar form except for the additional spin labels,
\begin{equation}
\mathbf{X}^{(u, u)}_{k^{\prime} \ell^{\prime} m^{\prime} \boldsymbol{m}_{s}^{\prime *}; k \ell m \boldsymbol{m}_{s}^*}\,,
\end{equation}
where we recall that the spin indices are defined using the dimer-axis convention, see \Cref{eq:dimeraxisms}. We now introduce the antisymmetrizing operation
\begin{multline}
\mathbf{X}\left(
\boldsymbol{k}^{\prime}, m_s(\kprimelittle); \boldsymbol{a}^{\prime}, m_s(\aprimelittle); \boldsymbol{b}^{\prime}, m_s(\bprimelittle) |
\boldsymbol{k}, m_s(\klittle); \boldsymbol{a}, m_s(\alittle); \boldsymbol{b}, m_s(\blittle)\right) \equiv
\\
\cA\left[
\mathbf{X}^{(u, u)}_{k^{\prime} \ell^{\prime} m^{\prime} \boldsymbol{m}_{s}^{\prime *}; k \ell m \boldsymbol{m}_{s}^*}
\right]\,,
\label{eq:antisymmetrizedef}
\end{multline}
which is defined by first combining with spherical harmonics and converting spin indices to the lab-axis frame [see \Cref{eq:spin_short}],
\begin{multline}
\mathbf{X}^{(u, u)}\left(
\boldsymbol{k}^{\prime}, m_s(\kprimelittle);
\boldsymbol{a}^{\prime}, m_s(\aprimelittle);
\boldsymbol{b}^{\prime}, m_s(\bprimelittle) |
\boldsymbol{k}, m_s(\klittle);
\boldsymbol{a}, m_s(\alittle);
\boldsymbol{b}, m_s(\blittle)\right)
\equiv
\\
4 \pi Y_{\ell^{\prime} m^{\prime}}^{*}(\hat{\boldsymbol{a}}_{2, k^{\prime}}^{\prime *})
\mathcal{D}^{(k',a')}_{\bm m_s^{\prime} \bm m_s^{\prime *}}
\mathbf{X}^{(u, u)}_{k^{\prime} \ell^{\prime} m^{\prime} \boldsymbol{m}_{s}^{\prime*} ; k \ell m \boldsymbol{m}^*_{s}}
\mathcal{D}^{(k,a)\dagger}_{\bm m_s^* \bm m_s}
Y_{\ell m}(\hat{\boldsymbol{a}}_{2, k}^{*}) \,,
\end{multline}
and then antisymmetrizing by using the following sum,
\begin{multline}
\mathbf{X}\left(
\boldsymbol{k}^{\prime}, m_s(\kprimelittle); \boldsymbol{a}^{\prime}, m_s(\aprimelittle); \boldsymbol{b}^{\prime}, m_s(\bprimelittle) |
\boldsymbol{k}, m_s(\bm k); \boldsymbol{a}, m_s(\bm a); \boldsymbol{b}, m_s(\bm b)\right) =
\\
\sum_{\left\{\boldsymbol{p}_{i}^{\prime}, m_{si}^{\prime}\right\} \in \mathcal{P}_{3}^{\prime}}
\sum_{\left\{\boldsymbol{p}_{i}, m_{si}\right\} \in \mathcal{P}_{3}}
\mathbf{X}^{(u, u)}\left(\boldsymbol{p}_1^{\prime}, m_{s 1}^{\prime}; \boldsymbol{p}_2^{\prime}, m_{s 2}^{\prime};
\boldsymbol{p}_2^{\prime}, m_{s 3}^{\prime} |
\boldsymbol{p}_1, m_{s 1}; \boldsymbol{p}_2, m_{s 2}; \boldsymbol{p}_3, m_{s 3}\right)\,.
\end{multline}
The permutations involve changing both momenta and the corresponding spin components,
\begin{equation}
\mathcal{P}_{3}=\{\{\boldsymbol{k}, m_{s1}, \boldsymbol{a}, m_{s2}, \boldsymbol{b}, m_{s3} \},\{ \boldsymbol{a}, m_{s2}, \boldsymbol{b}, m_{s3}, \boldsymbol{k}, m_{s1} \},\{\boldsymbol{b}, m_{s3}, \boldsymbol{k}, m_{s1}, \boldsymbol{a}, m_{s2} \}\}\,,
\end{equation}
and similarly for $\mathcal{P}_{3}^\prime$. Here we are using the shorthand
\begin{equation}
m_{s1}=m(\klittle)\,,\quad
m_{s2}=m(\alittle)\,,\quad
m_{s3}=m(\blittle)\,,
\end{equation}
and similarly for the primed spin indices.

Note that antisymmetrization only requires adding terms with a positive sign, since we are using cyclic permutations. Antisymmetry under single permutations is built in to the matrix version of $\textbf X^{(u,u)}$.

\section{Operators with two derivatives\label{app:twoderivs}}

In this appendix we determine all the Lorentz- and parity-invariant operators of the form
\begin{equation}
``\partial^2" (\overline{\mathcal N}\Gamma_1 {\mathcal N}) (\overline{\mathcal N}\Gamma_2 {\mathcal N}) (\overline{\mathcal N}\Gamma_3 {\mathcal N})\,.
\label{eq:general2deriv}
\end{equation}
Here $\Gamma_i$ are Dirac matrices, and the derivatives can act on any of the fields. We allow the use of the noninteracting equations of motion to reduce the set of operators, which is equivalent to considering on-shell matrix elements. We also assume that the total momentum inserted by the operator vanishes, so that we can freely integrate by parts.

We now argue that the general operator in \Cref{eq:general2deriv} can be brought into a canonical form
\begin{equation}
([\partial\overline{\mathcal N}]\Gamma_1 [\partial {\mathcal N}]) (\overline{\mathcal N}\Gamma_2 {\mathcal N}) (\overline{\mathcal N}\Gamma_3 {\mathcal N})\,,
\label{eq:canonical2deriv}
\end{equation}
in which both derivatives act within the same bilinear, one on $\overline{\mathcal N}$ and the other on ${\mathcal N}$. The two derivatives in the general operators can act in three ways: (i) both on factors of $\overline{\mathcal N}$, (ii) both on factors of ${\mathcal N}$, or (iii) one on each. In cases (i) and (ii), if both act on the same field, we can use the equations of motion to remove the derivatives: $\partial^2 \to -m^2$. In case (i) if both act on different factors of $\overline{\mathcal N}$, we can integrate by parts and move one of the derivatives onto the other fields, yielding one term that involves a $\partial^2$, another that has two derivatives on different $\overline{\mathcal N}$ fields, and three terms with one derivative each on a $\overline{\mathcal N}$ and a ${\mathcal N}$. The term with derivatives on two different $\overline{\mathcal N}$ fields can be moved to the left-hand side of the (now matrix) equation, while, using Fierz identities, we can bring terms with derivatives on $\overline{\mathcal N}$ and ${\mathcal N}$ fields in different bilinears to a form in which both the derivatives act within the same bilinear. Assuming that we can invert this matrix equation, each term with derivatives on different factors of $\overline{\mathcal N}$ can be rewritten in our canonical form together with a nonderivative term. An essentially identical argument holds for case (ii), with the roles of $\overline{\mathcal N}$ and ${\mathcal N}$ interchanged. For case (iii), if needed we can use Fierz identities to reach the canonical form.

We find 30 operators of the canoncial form to be independent, 8 of which can be dropped using the equations of motion. For convenience, we divide these into those in which the Lorentz indices on the derivatives are contracted together (ten in all)%
\footnote{%
In all operators, derivatives act only on the object immediately to their right.
}
\begin{align}
SSS &= (\partial^\mu \overline{\mathcal N} \partial_\mu {\mathcal N}) (\overline{\mathcal N} {\mathcal N}) (\overline{\mathcal N} {\mathcal N})\,,
\\
SPP &=(\partial^\mu \overline{\mathcal N} \partial_\mu {\mathcal N}) (\overline{\mathcal N}\gamma_5 {\mathcal N}) (\overline{\mathcal N} \gamma_5 {\mathcal N})\,,
\\
PSP &=(\partial^\mu \overline{\mathcal N} \gamma_5\partial_\mu {\mathcal N}) (\overline{\mathcal N} {\mathcal N}) (\overline{\mathcal N} \gamma_5 {\mathcal N})\,,
\\
SVV &=(\partial^\mu \overline{\mathcal N} \partial_\mu {\mathcal N}) (\overline{\mathcal N}\gamma_\nu {\mathcal N}) (\overline{\mathcal N}\gamma^\nu {\mathcal N})\,,
\\
VSV &=(\partial^\mu \overline{\mathcal N} \gamma_\nu \partial_\mu {\mathcal N}) (\overline{\mathcal N} {\mathcal N}) (\overline{\mathcal N}\gamma^\nu {\mathcal N})\,,
\\
ASA &=(\partial^\mu \overline{\mathcal N} \gamma_\nu\gamma_5 \partial_\mu {\mathcal N}) (\overline{\mathcal N} {\mathcal N})
(\overline{\mathcal N}\gamma^\nu \gamma_5 {\mathcal N})\,,
\\
TST &=(\partial^\mu \overline{\mathcal N} \sigma_{\nu\rho} \partial_\mu {\mathcal N}) (\overline{\mathcal N} {\mathcal N}) (\overline{\mathcal N}\sigma^{\nu\rho} {\mathcal N})\,,
\\
PVA &=(\partial^\mu \overline{\mathcal N} \gamma_5 \partial_\mu {\mathcal N}) (\overline{\mathcal N} \gamma_\nu {\mathcal N})
(\overline{\mathcal N}\gamma^\nu \gamma_5 {\mathcal N})\,,
\\
VAP &=(\partial^\mu \overline{\mathcal N} \gamma^\nu \partial_\mu {\mathcal N}) (\overline{\mathcal N} \gamma_\nu \gamma_5 {\mathcal N})
(\overline{\mathcal N} \gamma_5 {\mathcal N})\,,
\\
APV &=(\partial^\mu \overline{\mathcal N} \gamma^\nu\gamma_5 \partial_\mu {\mathcal N}) (\overline{\mathcal N} \gamma_5 {\mathcal N})
(\overline{\mathcal N} \gamma_\nu {\mathcal N})\,,
\end{align}
those in which the derivatives are contracted with a Dirac matrix (12 in all),
\begin{align}
SVV' &=(\partial^\mu \overline{\mathcal N} \partial_\nu {\mathcal N}) (\overline{\mathcal N}\gamma_\mu {\mathcal N}) (\overline{\mathcal N}\gamma^\nu {\mathcal N})\,,
\\
TST' &=(\partial_\nu \overline{\mathcal N} \sigma_{\mu\rho} \partial^\mu {\mathcal N}) (\overline{\mathcal N} {\mathcal N}) (\overline{\mathcal N}\sigma^{\nu\rho} {\mathcal N})\,,
\\
PVA' &=(\partial^\mu \overline{\mathcal N} \gamma_5 \partial_\nu {\mathcal N}) (\overline{\mathcal N} \gamma_\mu {\mathcal N})
(\overline{\mathcal N}\gamma^\nu \gamma_5 {\mathcal N})\,,
\\
VTV' &=(\partial^\mu \overline{\mathcal N} \gamma_\rho \partial_\nu {\mathcal N}) (\overline{\mathcal N} \sigma^{\nu\rho} {\mathcal N})
(\overline{\mathcal N}\gamma_\mu {\mathcal N})\,,
\\
ATA' &=(\partial^\mu \overline{\mathcal N} \gamma_\rho\gamma_5 \partial_\nu {\mathcal N}) (\overline{\mathcal N} \sigma^{\nu\rho} {\mathcal N})
(\overline{\mathcal N}\gamma_\mu \gamma_5 {\mathcal N})\,,
\\
TTT' &=(\partial^\mu \overline{\mathcal N} \sigma_{\mu\eta} \partial^\eta {\mathcal N}) (\overline{\mathcal N} \sigma^{\nu\rho} {\mathcal N})
(\overline{\mathcal N}\sigma_{\nu\rho} {\mathcal N})\,,
\\
TTT'' &=(\partial_\mu \overline{\mathcal N} \sigma^{\mu\nu} \partial_\eta {\mathcal N}) (\overline{\mathcal N} \sigma^{\eta\rho} {\mathcal N})
(\overline{\mathcal N}\sigma_{\rho\nu} {\mathcal N})\,,
\\
TTT'' &=(\partial_\eta \overline{\mathcal N} \sigma^{\mu\nu} \partial_\mu {\mathcal N}) (\overline{\mathcal N} \sigma^{\eta\rho} {\mathcal N})
(\overline{\mathcal N}\sigma_{\rho\nu} {\mathcal N})\,,
\\
TT_5P' &=(\partial^\mu \overline{\mathcal N} \sigma_{\mu\rho } \partial_\nu {\mathcal N}) (\overline{\mathcal N} \sigma^{\nu\rho} \gamma_5{\mathcal N})
(\overline{\mathcal N} \gamma_5 {\mathcal N})\,,
\\
T_5VA' &=(\partial^\mu \overline{\mathcal N} \sigma_{\mu\nu} \gamma_5\partial^\nu {\mathcal N}) (\overline{\mathcal N} \gamma^{\rho} {\mathcal N})
(\overline{\mathcal N}\gamma_\rho\gamma_5 {\mathcal N})\,,
\\
T_5VA'' &=(\partial^\mu \overline{\mathcal N} \sigma_{\rho\nu} \gamma_5\partial^\nu {\mathcal N}) (\overline{\mathcal N} \gamma^{\mu} {\mathcal N})
(\overline{\mathcal N}\gamma^\rho\gamma_5 {\mathcal N})\,,
\\
T_5TT_5' &=(\partial^\mu \overline{\mathcal N} \sigma_{\mu\nu}\gamma_5 \partial^\nu {\mathcal N}) (\overline{\mathcal N} \sigma^{\eta\rho} {\mathcal N})
(\overline{\mathcal N}\sigma_{\eta\rho}\gamma_5 {\mathcal N})
\end{align}
and those that can be dropped using the equations of motion (8 in all)
\begin{align}
VSV' &=(\partial^\mu \overline{\mathcal N} \gamma_\mu \partial^\nu {\mathcal N}) (\overline{\mathcal N} {\mathcal N}) (\overline{\mathcal N}\gamma_\nu {\mathcal N})\,,
\\
VSV'' &=(\partial^\nu \overline{\mathcal N} \gamma_\mu \partial^\mu {\mathcal N}) (\overline{\mathcal N} {\mathcal N}) (\overline{\mathcal N}\gamma_\nu {\mathcal N})\,,
\\
ASA' &=(\partial^\mu \overline{\mathcal N} \gamma_\mu\gamma_5 \partial^\nu {\mathcal N}) (\overline{\mathcal N} {\mathcal N})
(\overline{\mathcal N}\gamma_\nu \gamma_5 {\mathcal N})\,,
\\
ASA'' &=(\partial^\nu \overline{\mathcal N} \gamma_\mu\gamma_5 \partial^\mu {\mathcal N}) (\overline{\mathcal N} {\mathcal N})
(\overline{\mathcal N}\gamma_\nu \gamma_5 {\mathcal N})\,,
\\
VAP' &=(\partial^\mu \overline{\mathcal N} \gamma_\mu \partial^\nu {\mathcal N}) (\overline{\mathcal N} \gamma_\nu \gamma_5 {\mathcal N})
(\overline{\mathcal N} \gamma_5 {\mathcal N})\,,
\\
VAP'' &=(\partial^\nu \overline{\mathcal N} \gamma_\mu \partial^\mu {\mathcal N}) (\overline{\mathcal N} \gamma_\nu \gamma_5 {\mathcal N})
(\overline{\mathcal N} \gamma_5 {\mathcal N})\,,
\\
APV' &=(\partial^\mu \overline{\mathcal N} \gamma_\mu\gamma_5 \partial^\nu {\mathcal N}) (\overline{\mathcal N} \gamma_5 {\mathcal N})
(\overline{\mathcal N} \gamma_\nu {\mathcal N})\,,
\\
APV'' &=(\partial^\nu \overline{\mathcal N} \gamma_\mu\gamma_5 \partial^\mu {\mathcal N}) (\overline{\mathcal N} \gamma_5 {\mathcal N})
(\overline{\mathcal N} \gamma_\nu {\mathcal N})
\,.
\end{align}

We have checked this result using an alternative method in which we explicitly include every possible position of the action of the two derivatives, and all possible contractions with Dirac matrices, without using the analytic simplifications of this appendix. We again find that only two operators appear at quadratic order.

\bibliographystyle{JHEP}
\bibliography{ref.bib}

\end{document}